\documentclass{article}

    \PassOptionsToPackage{numbers, sort, compress}{natbib}

\usepackage[final]{neurips_2023}

\usepackage[utf8]{inputenc} 
\usepackage[T1]{fontenc}    
\usepackage{hyperref}       
\usepackage{url}            
\usepackage{booktabs}       
\usepackage{amsfonts}       
\usepackage{nicefrac}       
\usepackage{microtype}      
\usepackage{xcolor}         
\usepackage{graphicx}

\usepackage{tabu}
\usepackage{multirow}
\usepackage{float}
\usepackage{rotating}
\usepackage{algpseudocode}
\usepackage{microtype}
\usepackage{subcaption}
\usepackage{algorithm}

\usepackage{amsmath}
\usepackage{amssymb}
\usepackage{mathtools}
\usepackage{amsthm}

\theoremstyle{plain}

\theoremstyle{definition}

\theoremstyle{remark}

\title{Generative Antibody Design for Complementary Chain Pairing Sequences through Encoder-Decoder Language Model}

\author{%
  Simon K.S.~Chu\thanks{Work done as an intern at Therapeutic Discovery, Amgen Research, Amgen Inc.} \\
  University of California Davis \\
  Davis, CA 95616 \\
  \texttt{kschu@ucdavis.edu} \\
  \And
  Kathy Y. ~Wei\thanks{Currently CSO and cofounder at 310 AI Inc.} \\
  Therapeutic Discovery, Amgen Research, Amgen Inc. \\
  South San Francisco, CA 94080 \\
  \texttt{kywei@alumni.stanford.edu} \\
}

\begin{document}

\maketitle

\begin{abstract}
Current protein language models (pLMs) predominantly focus on single-chain protein sequences and often have not accounted for constraints on generative design imposed by protein-protein interactions. To address this gap, we present paired Antibody T5 (pAbT5), an encoder-decoder model to generate complementary heavy or light chain from its pairing partner. We show that our model respects conservation in framework regions and variability in hypervariable domains, demonstrated by agreement with sequence alignment and variable-length CDR loops. We also show that our model captures chain pairing preferences through the recovery of ground-truth chain type and gene families. Our results showcase the potential of pAbT5 in generative antibody design, incorporating biological constraints from chain pairing preferences.
\end{abstract}

\section{Introduction}

Transformer-based protein language models (pLMs) have begun to find utility across a range of applications in the field. Remarkably, even when pretrained solely on sequence databases, these models have demonstrated the ability to aid in protein structure prediction \cite{Lin2022LanguagePrediction,Wu2022High-resolutionSequence} and a host of downstream tasks including function and secondary structure annotations \cite{Alley2019UnifiedLearning,Rao2019EvaluatingTAPE,Elnaggar2021ProtTrans:Learning,Brandes2022ProteinBERT:Function,Bachas2022AntibodyNaturalness}. Furthermore, they have shown promise in the area of \textit{de novo} protein design, proving to be useful in efforts ranging from point mutation design to full-sequence generation \cite{Rives2019BiologicalSequences,Ingraham2019GenerativeDesign,Madani2020ProGen:Generation,Meier2021LanguageFunction,Gligorijevic2021Function-guidedSampling,Nijkamp2022ProGen2:Models,Frey2023ProteinSampling}. By leveraging the evolutionary information contained in sequence databases, pLMs offer a pathway to understanding and designing protein sequences through a language modeling approach.

Most pLMs are designed for single-chain sequences only. However, many biological contexts involve protein-protein interactions where multiple chains interact simultaneously. For instance, antibodies consist of paired heavy and light chains. Modeling heavy and light chains independently is inadequate to reflect their heterodimeric nature and sacrifices their co-evolutionary information. Understanding antibody chain pairing has the potential to generate partner sequences given an existing heavy or light chain target.

To address this gap, we present paired Antibody T5 (pAbT5) to generate antibody sequences conditioned on their chain pairing partner in an encoder-decoder architecture. To summarize,
\begin{itemize}
\item {We modeled antibody chain pairing as a conditional protein design problem through T5 architecture.}
\item {We show that our model generates antibody sequences respecting conservation in framework region and variability in hypervariable domains.}
\item {We show that our generated sequences capture chain pairing preferences through the recovery of ground-truth chain type and gene families.}
\end{itemize}

\section{Related Work}
Prior works in generative antibody language models usually are based on either causal language models or denoising neural networks. \citet{Nijkamp2022ProGen2:Models} built a decoder-only model on single-chain antibody sequences. \citet{Shuai2022GenerativeDesign} extended the framework to conditional generation with species and chain type prefix tokens. Denoising network from \citet{Frey2023ProteinSampling} generates variable-length paired antibody sequences by introducing gap tokens. Distinct from language models, inverse folding models are capable of generating multiple-chain sequences based on structural inputs \cite{Dauparas2022RobustProteinMPNN,Hsu2022LearningStructures}.

\section{Methods}

\subsection{Model and Optimization}

We approach the antibody chain pairing problem under a sequence-to-sequence generation framework. We use the term forward-translation to describe light-to-heavy-chain generation and back-translation for the reciprocal process. Notably, we do not specify the translation direction, nor do we include any gap or prefix tokens relating to the input or target chain type, species, or gene families in our model. The model is fine-tuned from ProtT5-XL-UniRef50, which has a T5 architecture \cite{Elnaggar2021ProtTrans:Learning}.

To optimize our model, we adhere to the ProtT5-XL pretraining scheme utilizing a local batch size of 8 and a global batch size of 2048. We kept the encoder weights frozen and fine-tuned only on the decoder and observed better encoder representation on sequences compared to fine-tuning the whole model. We used a learning rate of 5e-5 without weight decay in AdaFactor optimizer with a gradient clipping of 1 and a patience of 5 epochs on validation loss for two days on eight A100 GPUs. The implementation is on PyTorch under HuggingFace framework \cite{Paszke2019PyTorch:Library,Wolf2020Transformers:Processing}.

\subsection{Dataset}

We sourced approximately 160k pairs of antibody VH and VL sequences from the Observed Antibody Space (OAS) database \cite{Olsen2022ObservedSequences}. Leveraging the framework of forward- and back-translations, we represented each bi-directional pairing through two uni-direction translations. This yielded a dataset of roughly 321k translation samples derived from 239k distinct sequences from humans, rats, and mice.

In the context of the protein-protein interaction network in the OAS dataset, edge-based splitting serves as an intuitive method for data partitioning. An alternative approach is node-based partitioning, where all edges linked to training nodes are incorporated into the training set, leaving the rest for testing. We employ an exclusive node split strategy, reserving specific nodes and their related edges solely for testing to rigorously evaluate the model's generalization to unseen sequences and pairings (Figure \ref{fig:graph_split}). Consequently, our dataset is partitioned into a roughly 90-5-5 distribution, resulting in 260k training, 828 validation, and 802 test translations.



\section{Results}

\subsection{Sequence Generation Aligns with Conserved and Variable Domains in Antibodies}
Antibodies display significant diversity in their hypervariable domains to ensure specificity in antigen binding. Both the light and heavy chains possess three loop structures, known as the CDR loops. While these loops are highly variable, other regions, termed framework regions, remain relatively conserved. Of all the CDR loops, the third loop on the heavy chain (CDRH3) exhibits the highest variability. In this section, we evaluate whether our model successfully recognizes and reproduces these distinct patterns during next-word prediction and sequence generation.

\begin{figure*}[h]
  \centering
  \includegraphics[width=\linewidth]{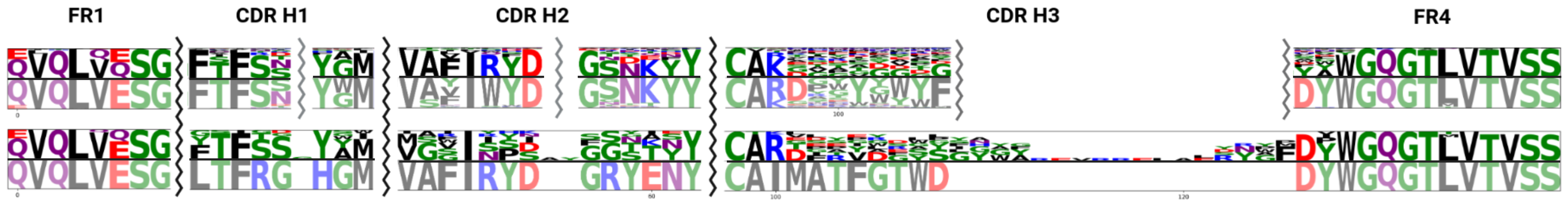}
  \caption{Comparison of observed and model-derived alignment profiles on heavy chain across framework regions (FR) and CDR loops. The first and second rows pair the next-word probability under teacher-forcing with sequence conservation from alignment to UniRef90 \cite{Suzek2007UniRef:Clusters}. The third and fourth rows provide a side-by-side view of global alignments between generated sequences and their corresponding observed sequences. The reciprocal analysis on the reverse direction can be found in Figure \ref{fig:alignment_profile_light}.}
  \label{fig:alignment_profile}
\end{figure*}

In Figure \ref{fig:alignment_profile}, we compare the probability from next-word prediction against conservation from alignment analysis. The model demonstrates higher confidence in the more conserved framework region of the heavy chain target and displays increased uncertainty in the variable CDR loops. To further examine its ability to generate realistic sequences, we align the observed and generated sequences for a random heavy-light chain pairing from the test set. Notably, generated sequences often exhibit greater variability than next-word probabilities, potentially due to the cascading effect during iterative sampling. These sequences might also originate from different gene loci or families than the target sequences. This analysis highlights the model's ability to generate variable-length CDR (H3) loops while preserving patterns in framework regions. On average, generated sequences maintain approximately 60\% whole-sequence identity with target sequences. This suggests our model effectively balances capturing antibody pairing patterns and creating novel sequences. For a detailed analysis of sequence identities and lengths by region, see Tables \ref{tab:generation_identity_regions}, \ref{tab:generation_length_regions} and \ref{tab:generation_identity}. Comprehensive alignment profiles for both heavy and light chains, along with four other random output samples from the test set, can be found in Figures \ref{fig:alignment_profile_light}, \ref{fig:alignment_profile_full}, \ref{fig:alignment_profile_full_light}, and \ref{fig:alignment_profile_examples}.

\begin{figure}[h]
  \centering
  \includegraphics[width=\linewidth]{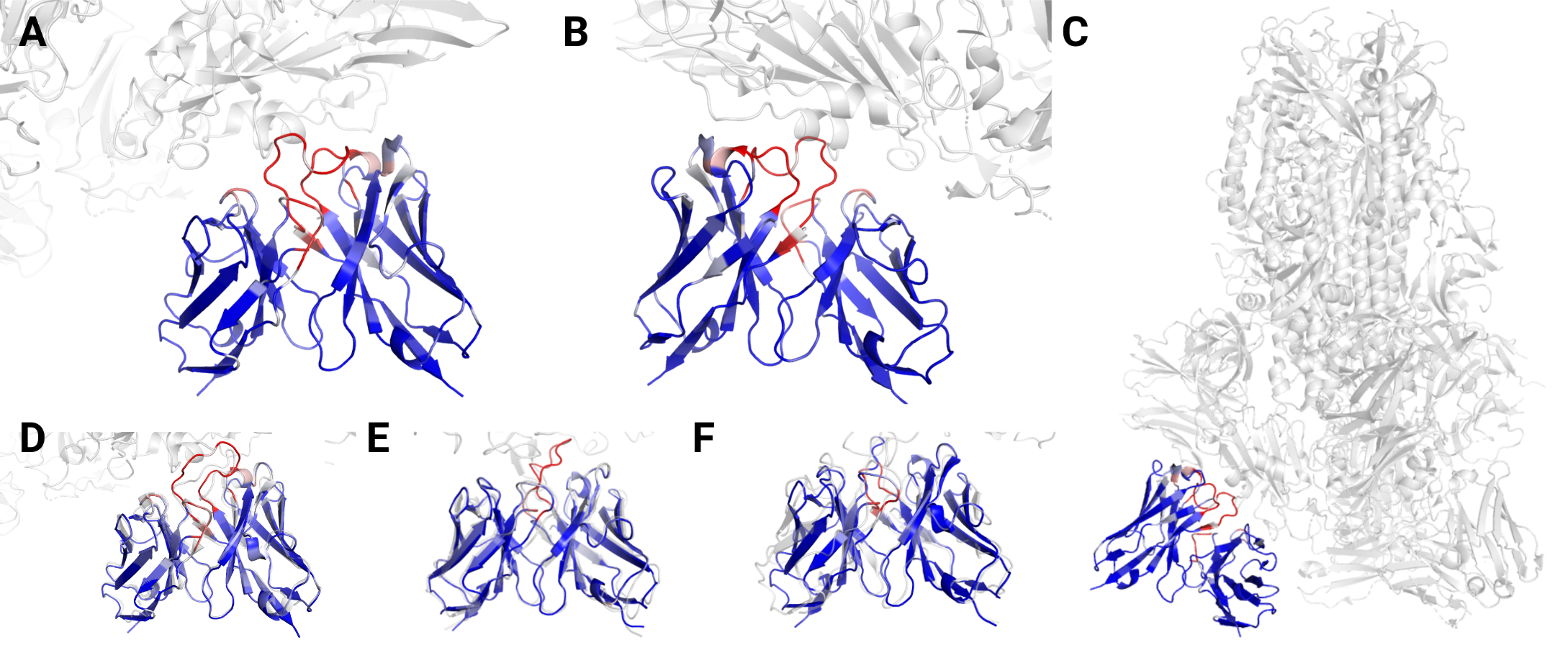}
  \caption{Visualization of next-word prediction entropy for antibodies bound to the SARS-CoV-2 spike protein. Blue denotes low entropy regions, while red represents areas of high entropy. (A and B) Offer front and back views of the 6WPS PDB structure. (C) Provides an overview of the entire 6WPS structure. (D, E, and F) depict structures from 6WPT, 7TB8 chains D and E, and 7TB8 chains H and I, respectively.}
  \label{fig:covid}
\end{figure}

Beyond assessing alignment profiles, we further validate our model predictions by superimposing these results on both predicted structures by DeepAb \cite{Ruffolo2022AntibodyLearning} and known experimental structures. Indeed, the generated heavy and light chains exhibit structurally consistent framework regions while emphasizing variations in the CDR loops, as illustrated in Figure \ref{fig:generated_structures}. With the interest to evaluate on unseen experimental structures, we analyzed three antibodies bound to the SARS-CoV-2 spike protein from RCSB database \cite{Pinto2020Cross-neutralizationAntibody,Zhou2022StructuralB.1.1.529}. Figure \ref{fig:covid} demonstrates that the CDR loops remain the most entropic regions across all three antibody structures.

\subsection{Conditional Generation Recovers Pairing Sequences}
\label{subsection:representation}

The human immune system can recognize a vast array of antigens by generating a diverse repertoire of antibodies through gene rearrangement. The transcription of each antibody sequence is driven by the combination of C, V, and J genes, with an additional D gene specifically for the heavy chain. These genes are stored within chromosome gene loci, specifically H, $\lambda$, and $\kappa$, and each of these genes corresponds to a segment within the complete antibody sequence. The recombination of VDJ gene families allows for an impressive array of heavy-light chain pairings, estimated at around $10^6$ combinations, which are further amplified by somatic mutations \cite{berg2007biochemistry}

To benchmark our generative model against the current state-of-the-art, we evaluate the percentage of generated sequences sharing the same chain type, gene loci, V, and J gene families as the pairing target. We assess our model against ProGen2-OAS and IgLM, two publicly available state-of-the-art antibody language models. ProGen2-OAS is a decoder-only LM trained on unpaired antibody sequences, and as such, it's not inherently designed to understand antibody pairing \cite{Nijkamp2022ProGen2:Models}. Similarly, IgLM, while focusing on conditions of species and chain type by appending tag tokens at sequence starts, doesn't have an inherent design for pairing comprehension \cite{Shuai2022GenerativeDesign}. For ProGen2-OAS, pairing sequences are generated unconditionally. In the IgLM scenario, we provide chain and species tags, assuming the heavy chain must pair with the light chain and both chains belong to the same species. Additionally, we introduce a baseline of selecting a sequence at random from the test set population and another baseline of selecting the pairing partner of the closest sequence from the validation set, termed as \textit{population sampling} and \textit{closest sequence}.

In Table \ref{tab:recovery_rate}, we present a comparison of the percentage of generated sequences that align with the target across various attributes. pAbT5 consistently demonstrates superior performance compared to current state-of-the-art models and baselines. Our model's efficacy significantly surpasses that of \textit{population sampling} and  \textit{closest sequence}, suggesting that pAbT5's target recovery is not merely from exploiting dataset biases or memorization. We highlight the importance of the encoder in the T5 architecture by removing cross-attention and retraining on the decoder-only model, which results in a similar performance to \textit{population sampling}. It's notable that ProGen2-OAS exhibits a marked preference for generating heavy chain sequences, aligning with observations from the unpaired OAS dataset \cite{Olsen2022ObservedSequences}. \citet{Nijkamp2022ProGen2:Models} assessed their model by starting sequence generation with the first few tokens. Contrarily, we decided against providing these initial tokens to ensure no possible clues about target gene loci or families were given, especially when these details aren't evident from the chain type of the pairing partner alone. Even when provided with chain type and species tags, IgLM doesn't quite match the performance of our model. pAbT5 sets a new benchmark in most areas, with the exception being gene families with smaller sample sizes, as illustrated in Figure \ref{fig:generation_baseline} and \ref{fig:generation_sunbursts}.

\begin{table*}[h]
  \centering
      \begin{tabular}{c c c c c}
        \multicolumn{5}{c}{\textbf{Percentage of generated sequences sharing the same attributes with target}} \\
        \hline
        & Chain type & Gene loci & V gene family & J gene family \\
        \hline
        Population sampling & 0.50 (2163) & 0.38 (1644) & 0.12 (513) & 0.09 (394) \\
        Closest sequence & \textbf{1.00} (4300) & \textbf{0.76} (3290) & 0.17 (750) & 0.04 (180) \\ 
        ProGen2-OAS & 0.50 (2171) & 0.50 (2140) & 0.12 (500) & 0.01 (50) \\
        IgLM & \textbf{1.00} (4320) & \textbf{0.76} (3280) & 0.18 (763) & 0.03 (133) \\
        Our method (decoder-only) & 0.50 (2165) & 0.35 (1506) & 0.07 (319) & 0.09 (400) \\
        Our method (pAbT5) & \textbf{1.00} (4320) & \textbf{0.78} (3373) & \textbf{0.25} (1066) & \textbf{0.21} (896) \\
        \hline
        \end{tabular}
  \caption{Percentage of generated sequences in the human antibody test set that match the target sequence's chain type, gene loci, and V and J gene families. We compare our model with a random sequence from the population (population sampling), the pairing partner of the nearest sequence from the validation set (closest sequence), ProGen2-OAS \cite{Nijkamp2022ProGen2:Models}, IgLM \cite{Shuai2022GenerativeDesign}, and a decoder-only T5 model by removing the encoder and cross-attention.}
  \label{tab:recovery_rate}
\end{table*}

\section{Conclusion}
In this study, we introduced and evaluated pAbT5, demonstrating its efficacy in capturing intricate antibody pairing patterns and generating chain pairing sequences with notable precision relative to target attributes. Its performance, when compared to existing models, suggests its utility as a valuable tool in advancing antibody research and therapeutic exploration.

\section{Acknowledgements}  
We give our special thanks to Ai Ching Lim and Christy Tinberg for their generous support of this project. We thank George Seegan for language model discussion. We thank Yi Zheng, Danyang Gong, and Austin Rice for helpful discussion on gene families and applications in antibodies. We thank Grant Keller for introducing ANARCI.

\bibliographystyle{unsrtnat}  
\bibliography{references,methods}

\begin{thebibliography}{36}
\providecommand{\natexlab}[1]{#1}
\providecommand{\url}[1]{\texttt{#1}}
\expandafter\ifx\csname urlstyle\endcsname\relax
  \providecommand{\doi}[1]{doi: #1}\else
  \providecommand{\doi}{doi: \begingroup \urlstyle{rm}\Url}\fi

\bibitem[Lin et~al.(2022)Lin, Akin, Rao, Hie, Zhu, Lu, Dos, Costa, Fazel-Zarandi, Sercu, Candido, Rives, and Ai]{Lin2022LanguagePrediction}
Zeming Lin, Halil Akin, Roshan Rao, Brian Hie, Zhongkai Zhu, Wenting Lu, Allan Dos, Santos Costa, Maryam Fazel-Zarandi, Tom Sercu, Sal Candido, Alexander Rives, and Meta Ai.
\newblock {Language models of protein sequences at the scale of evolution enable accurate structure prediction}.
\newblock \emph{bioRxiv}, 2022.
\newblock \doi{10.1101/2022.07.20.500902}.
\newblock URL \url{https://www.biorxiv.org/content/early/2022/10/31/2022.07.20.500902}.

\bibitem[Wu et~al.(2022)Wu, Ding, Wang, Shen, Zhang, Luo, Su, Wu, Xie, Berger, Ma, and Peng]{Wu2022High-resolutionSequence}
Ruidong Wu, Fan Ding, Rui Wang, Rui Shen, Xiwen Zhang, Shitong Luo, Chenpeng Su, Zuofan Wu, Qi~Xie, Bonnie Berger, Jianzhu Ma, and Jian Peng.
\newblock {High-resolution de novo structure prediction from primary sequence}.
\newblock \emph{bioRxiv}, 2022.
\newblock \doi{10.1101/2022.07.21.500999}.
\newblock URL \url{http://biorxiv.org/content/early/2022/07/22/2022.07.21.500999.abstract}.

\bibitem[Alley et~al.(2019)Alley, Khimulya, Biswas, AlQuraishi, and Church]{Alley2019UnifiedLearning}
Ethan~C. Alley, Grigory Khimulya, Surojit Biswas, Mohammed AlQuraishi, and George~M. Church.
\newblock {Unified rational protein engineering with sequence-based deep representation learning}.
\newblock \emph{Nature Methods}, 16\penalty0 (12):\penalty0 1315--1322, 12 2019.
\newblock ISSN 15487105.
\newblock \doi{10.1038/s41592-019-0598-1}.

\bibitem[Rao et~al.(2019)Rao, Bhattacharya, Thomas, Duan, Chen, Canny, Abbeel, and Song]{Rao2019EvaluatingTAPE}
Roshan Rao, Nicholas Bhattacharya, Neil Thomas, Yan Duan, Xi~Chen, John Canny, Pieter Abbeel, and Yun~S Song.
\newblock {Evaluating Protein Transfer Learning with TAPE}.
\newblock \emph{Advances in Neural Information Processing Systems}, 32, 2019.
\newblock URL \url{https://github.com/songlab-cal/tape}.

\bibitem[Elnaggar et~al.(2021)Elnaggar, Heinzinger, Dallago, Rehawi, Wang, Jones, Gibbs, Feher, Angerer, Steinegger, Bhowmik, and Rost]{Elnaggar2021ProtTrans:Learning}
Ahmed Elnaggar, Michael Heinzinger, Christian Dallago, Ghalia Rehawi, Yu~Wang, Llion Jones, Tom Gibbs, Tamas Feher, Christoph Angerer, Martin Steinegger, Debsindhu Bhowmik, and Burkhard Rost.
\newblock {ProtTrans: Towards Cracking the Language of Life's Code Through Self-Supervised Learning}.
\newblock \emph{bioRxiv}, 14\penalty0 (8), 2021.
\newblock URL \url{https://www.biorxiv.org/content/early/2021/05/04/2020.07.12.199554}.

\bibitem[Brandes et~al.(2022)Brandes, Ofer, Peleg, Rappoport, and Linial]{Brandes2022ProteinBERT:Function}
Nadav Brandes, Dan Ofer, Yam Peleg, Nadav Rappoport, and Michal Linial.
\newblock {ProteinBERT: a universal deep-learning model of protein sequence and function}.
\newblock \emph{Bioinformatics}, 38\penalty0 (8):\penalty0 2102--2110, 4 2022.
\newblock ISSN 14602059.
\newblock \doi{10.1093/bioinformatics/btac020}.

\bibitem[Bachas et~al.(2022)Bachas, Rakocevic, Spencer, Sastry, Haile, Sutton, Kasun, Stachyra, Gutierrez, Yassine, Medjo, Blay, Kohnert, Stanton, Brown, Tijanic, Mccloskey, Viazzo, Consbruck, Carter, Levine, Abdulhaqq, Shaul, Ventura, Olson, Yapici, Meier, Mcclain, Weinstock, Hannum, Schwartz, Gander, and Spreafico]{Bachas2022AntibodyNaturalness}
Sharrol Bachas, Goran Rakocevic, David Spencer, Anand~V Sastry, Robel Haile, John~M Sutton, George Kasun, Andrew Stachyra, Jahir~M Gutierrez, Edriss Yassine, Borka Medjo, Vincent Blay, Christa Kohnert, Jennifer~T Stanton, Alexander Brown, Nebojsa Tijanic, Cailen Mccloskey, Rebecca Viazzo, Rebecca Consbruck, Hayley Carter, Simon Levine, Shaheed Abdulhaqq, Jacob Shaul, Abigail~B Ventura, Randal~S Olson, Engin Yapici, Joshua Meier, Sean Mcclain, Matthew Weinstock, Gregory Hannum, Ariel Schwartz, Miles Gander, and Roberto Spreafico.
\newblock {Antibody optimization enabled by artificial intelligence predictions of binding affinity and naturalness}.
\newblock \emph{bioRxiv}, 2022.
\newblock \doi{10.1101/2022.08.16.504181}.
\newblock URL \url{https://www.biorxiv.org/content/early/2022/08/17/2022.08.16.504181}.

\bibitem[Rives et~al.(2019)Rives, Meier, Sercu, Goyal, Lin, Liu, Guo, Ott, Zitnick, Ma, Fergus, Meier, Guo, Ott, Zitnick, Ma, and Fergus]{Rives2019BiologicalSequences}
Alexander Rives, Joshua Meier, Tom Sercu, Siddharth Goyal, Zeming Lin, Jason Liu, Demi Guo, Myle Ott, C.~Lawrence Zitnick, Jerry Ma, Rob Fergus, Joshua Meier, Demi Guo, Myle Ott, C.~Lawrence Zitnick, Jerry Ma, and Rob Fergus.
\newblock {Biological structure and function emerge from scaling unsupervised learning to 250 million protein sequences}.
\newblock \emph{bioRxiv}, 118\penalty0 (15):\penalty0 e2016239118, 4 2019.
\newblock \doi{10.1101/622803}.
\newblock URL \url{http://www.pnas.org/content/118/15/e2016239118.abstract}.

\bibitem[Ingraham et~al.(2019)Ingraham, Garg, Barzilay, and Jaakkola]{Ingraham2019GenerativeDesign}
John Ingraham, Vikas~K Garg, Regina Barzilay, and Tommi Jaakkola.
\newblock {Generative models for graph-based protein design}.
\newblock \emph{Advances in Neural Information Processing Systems}, pages 15820--15831, 2019.

\bibitem[Madani et~al.(2020)Madani, McCann, Naik, Keskar, Anand, Eguchi, Huang, and Socher]{Madani2020ProGen:Generation}
Ali Madani, Bryan McCann, Nikhil Naik, Nitish~Shirish Keskar, Namrata Anand, Raphael~R. Eguchi, Po-Ssu Huang, and Richard Socher.
\newblock {ProGen: Language Modeling for Protein Generation}.
\newblock \emph{arXiv}, 3 2020.
\newblock URL \url{http://arxiv.org/abs/2004.03497}.

\bibitem[Meier et~al.(2021)Meier, Rao, Verkuil, Liu, Sercu, and Rives]{Meier2021LanguageFunction}
Joshua Meier, Roshan Rao, Robert Verkuil, Jason Liu, Tom Sercu, and Alexander Rives.
\newblock {Language models enable zero-shot prediction of the effects of mutations on protein function}.
\newblock \emph{Advances in Neural Information Processing Systems}, 34:\penalty0 1--28, 2021.
\newblock URL \url{https://proceedings.neurips.cc/paper/2021/hash/f51338d736f95dd42427296047067694-Abstract.html}.

\bibitem[Gligorijevi{\'{c}} et~al.(2021)Gligorijevi{\'{c}}, Berenberg, Ra, Watkins, Kelow, Cho, and Bonneau]{Gligorijevic2021Function-guidedSampling}
Vladimir Gligorijevi{\'{c}}, Daniel Berenberg, Stephen Ra, Andrew Watkins, Simon Kelow, Kyunghyun Cho, and Richard Bonneau.
\newblock {Function-guided protein design by deep manifold sampling}.
\newblock \emph{bioRxiv}, 2021.
\newblock \doi{10.1101/2021.12.22.473759}.
\newblock URL \url{https://doi.org/10.1101/2021.12.22.473759}.

\bibitem[Nijkamp et~al.(2022)Nijkamp, Ruffolo, Weinstein, Naik, and Madani]{Nijkamp2022ProGen2:Models}
Erik Nijkamp, Jeffrey Ruffolo, Eli~N. Weinstein, Nikhil Naik, and Ali Madani.
\newblock {ProGen2: Exploring the Boundaries of Protein Language Models}.
\newblock \emph{arXiv}, 6 2022.
\newblock URL \url{http://arxiv.org/abs/2206.13517}.

\bibitem[Frey et~al.(2023)Frey, Berenberg, Zadorozhny, Kleinhenz, Lafrance-Vanasse, Hotzel, Wu, Ra, Bonneau, Cho, Loukas, Gligorijevic, and Saremi]{Frey2023ProteinSampling}
Nathan~C. Frey, Daniel Berenberg, Karina Zadorozhny, Joseph Kleinhenz, Julien Lafrance-Vanasse, Isidro Hotzel, Yan Wu, Stephen Ra, Richard Bonneau, Kyunghyun Cho, Andreas Loukas, Vladimir Gligorijevic, and Saeed Saremi.
\newblock {Protein Discovery with Discrete Walk-Jump Sampling}.
\newblock 6 2023.
\newblock URL \url{http://arxiv.org/abs/2306.12360}.

\bibitem[Shuai et~al.(2022)Shuai, Ruffolo, and Gray]{Shuai2022GenerativeDesign}
Richard~W Shuai, Jeffrey~A Ruffolo, and Jeffrey~J Gray.
\newblock {Generative Language Modeling for Antibody Design}.
\newblock \emph{bioRxiv}, 2022.
\newblock \doi{10.1101/2021.12.13.472419}.
\newblock URL \url{https://doi.org/10.1101/2021.12.13.472419}.

\bibitem[Dauparas et~al.(2022)Dauparas, Anishchenko, Bennett, Bai, Ragotte, Milles, Wicky, Courbet, De~Haas, Bethel, Leung, Huddy, Pellock, Tischer, Chan, Koepnick, Nguyen, Kang, Sankaran, Bera, King, and Baker]{Dauparas2022RobustProteinMPNN}
J~Dauparas, I~Anishchenko, N~Bennett, H~Bai, R~J Ragotte, L~F Milles, B~I~M Wicky, A~Courbet, R~J De~Haas, N~Bethel, P~J~Y Leung, T~F Huddy, S~Pellock, D~Tischer, F~Chan, B~Koepnick, H~Nguyen, A~Kang, B~Sankaran, A~K Bera, N~P King, and D~Baker.
\newblock {Robust deep learning-based protein sequence design using ProteinMPNN}.
\newblock \emph{Science}, 378:\penalty0 49--56, 2022.
\newblock URL \url{https://www.science.org}.

\bibitem[Hsu et~al.(2022)Hsu, Verkuil, Liu, Lin, Hie, Sercu, Lerer, and Rives]{Hsu2022LearningStructures}
Chloe Hsu, Robert Verkuil, Jason Liu, Zeming Lin, Brian Hie, Tom Sercu, Adam Lerer, and Alexander Rives.
\newblock {Learning inverse folding from millions of predicted structures}.
\newblock \emph{bioRxiv}, 2022.
\newblock \doi{10.1101/2022.04.10.487779}.
\newblock URL \url{https://doi.org/10.1101/2022.04.10.487779}.

\bibitem[Paszke et~al.(2019)Paszke, Gross, Massa, Lerer, Bradbury~Google, Chanan, Killeen, Lin, Gimelshein, Antiga, Desmaison, Xamla, Yang, Devito, Raison~Nabla, Tejani, Chilamkurthy, Ai, Steiner, Facebook, Facebook, and Chintala]{Paszke2019PyTorch:Library}
Adam Paszke, Sam Gross, Francisco Massa, Adam Lerer, James Bradbury~Google, Gregory Chanan, Trevor Killeen, Zeming Lin, Natalia Gimelshein, Luca Antiga, Alban Desmaison, Andreas~Köpf Xamla, Edward Yang, Zach Devito, Martin Raison~Nabla, Alykhan Tejani, Sasank Chilamkurthy, Qure Ai, Benoit Steiner, Lu~Fang Facebook, Junjie~Bai Facebook, and Soumith Chintala.
\newblock {PyTorch: An Imperative Style, High-Performance Deep Learning Library}.
\newblock In \emph{Advances in Neural Information Processing Systems}, 2019.

\bibitem[Wolf et~al.(2020)Wolf, Debut, Sanh, Chaumond, Delangue, Moi, Cistac, Rault, Louf, Funtowicz, Davison, Shleifer, Von~Platen, Ma, Jernite, Plu, Xu, Le~Scao, Gugger, Drame, Lhoest, and Rush]{Wolf2020Transformers:Processing}
Thomas Wolf, Lysandre Debut, Victor Sanh, Julien Chaumond, Clement Delangue, Anthony Moi, Pierric Cistac, Tim Rault, Rémi Louf, Morgan Funtowicz, Joe Davison, Sam Shleifer, Patrick Von~Platen, Clara Ma, Yacine Jernite, Julien Plu, Canwen Xu, Teven Le~Scao, Sylvain Gugger, Mariama Drame, Quentin Lhoest, and Alexander~M Rush.
\newblock {Transformers: State-of-the-Art Natural Language Processing}.
\newblock In \emph{Proceedings of the 2020 Conference on Empirical Methods in Natural Language Processing: System Demonstrations}, pages 38--45. Association for Computational Linguistics, 2020.
\newblock URL \url{https://github.com/huggingface/}.

\bibitem[Olsen et~al.(2022)Olsen, Boyles, and Deane]{Olsen2022ObservedSequences}
Tobias~H. Olsen, Fergus Boyles, and Charlotte~M. Deane.
\newblock {Observed Antibody Space: A diverse database of cleaned, annotated, and translated unpaired and paired antibody sequences}.
\newblock \emph{Protein Science}, 31\penalty0 (1):\penalty0 141--146, 1 2022.
\newblock ISSN 1469896X.
\newblock \doi{10.1002/pro.4205}.

\bibitem[Suzek et~al.(2007)Suzek, Huang, McGarvey, Mazumder, and Wu]{Suzek2007UniRef:Clusters}
Baris~E. Suzek, Hongzhan Huang, Peter McGarvey, Raja Mazumder, and Cathy~H. Wu.
\newblock {UniRef: Comprehensive and non-redundant UniProt reference clusters}.
\newblock \emph{Bioinformatics}, 23\penalty0 (10):\penalty0 1282--1288, 5 2007.
\newblock ISSN 13674803.
\newblock \doi{10.1093/bioinformatics/btm098}.

\bibitem[Ruffolo et~al.(2022)Ruffolo, Sulam, and Gray]{Ruffolo2022AntibodyLearning}
Jeffrey~A. Ruffolo, Jeremias Sulam, and Jeffrey~J. Gray.
\newblock {Antibody structure prediction using interpretable deep learning}.
\newblock \emph{Patterns}, 3\penalty0 (2), 2 2022.
\newblock ISSN 26663899.
\newblock \doi{10.1016/j.patter.2021.100406}.

\bibitem[Pinto et~al.(2020)Pinto, Park, Beltramello, Walls, Tortorici, Bianchi, Jaconi, Culap, Zatta, De~Marco, Peter, Guarino, Spreafico, Cameroni, Case, Chen, Havenar-Daughton, Snell, Telenti, Virgin, Lanzavecchia, Diamond, Fink, Veesler, and Corti]{Pinto2020Cross-neutralizationAntibody}
Dora Pinto, Young~Jun Park, Martina Beltramello, Alexandra~C. Walls, M.~Alejandra Tortorici, Siro Bianchi, Stefano Jaconi, Katja Culap, Fabrizia Zatta, Anna De~Marco, Alessia Peter, Barbara Guarino, Roberto Spreafico, Elisabetta Cameroni, James~Brett Case, Rita~E. Chen, Colin Havenar-Daughton, Gyorgy Snell, Amalio Telenti, Herbert~W. Virgin, Antonio Lanzavecchia, Michael~S. Diamond, Katja Fink, David Veesler, and Davide Corti.
\newblock {Cross-neutralization of SARS-CoV-2 by a human monoclonal SARS-CoV antibody}.
\newblock \emph{Nature}, 583\penalty0 (7815):\penalty0 290--295, 7 2020.
\newblock ISSN 14764687.
\newblock \doi{10.1038/s41586-020-2349-y}.

\bibitem[Zhou et~al.(2022)Zhou, Wang, Misasi, Pegu, Zhang, Harris, Olia, Talana, Yang, Chen, Choe, Shi, Teng, Creanga, Jenkins, Leung, Liu, Stancofski, Stephens, Zhang, Tsybovsky, Graham, Mascola, Sullivan, and Kwong]{Zhou2022StructuralB.1.1.529}
Tongqing Zhou, Lingshu Wang, John Misasi, Amarendra Pegu, Yi~Zhang, Darcy~R. Harris, Adam~S. Olia, Chloe~Adrienna Talana, Eun~Sung Yang, Man Chen, Misook Choe, Wei Shi, I.~Ting Teng, Adrian Creanga, Claudia Jenkins, Kwanyee Leung, Tracy Liu, Erik Stephane~D. Stancofski, Tyler Stephens, Baoshan Zhang, Yaroslav Tsybovsky, Barney~S. Graham, John~R. Mascola, Nancy~J. Sullivan, and Peter~D. Kwong.
\newblock {Structural basis for potent antibody neutralization of SARS-CoV-2 variants including B.1.1.529}.
\newblock \emph{Science}, 376\penalty0 (6591), 4 2022.
\newblock ISSN 10959203.
\newblock \doi{10.1126/science.abn8897}.

\bibitem[Berg et~al.(2007)Berg, Tymoczko, and Stryer]{berg2007biochemistry}
Jeremy~M Berg, John~L Tymoczko, and Lubert Stryer.
\newblock \emph{Biochemistry (Loose-Leaf)}.
\newblock Macmillan, 2007.

\bibitem[Dunbar and Deane(2016)]{Dunbar2016ANARCI:Classification}
James Dunbar and Charlotte~M. Deane.
\newblock {ANARCI: Antigen receptor numbering and receptor classification}.
\newblock \emph{Bioinformatics}, 32\penalty0 (2):\penalty0 298--300, 1 2016.
\newblock ISSN 14602059.
\newblock \doi{10.1093/bioinformatics/btv552}.

\bibitem[Larkin et~al.(2007)Larkin, Blackshields, Brown, Chenna, Mcgettigan, McWilliam, Valentin, Wallace, Wilm, Lopez, Thompson, Gibson, and Higgins]{Larkin2007Clustal2.0}
M.~A. Larkin, G.~Blackshields, N.~P. Brown, R.~Chenna, P.~A. Mcgettigan, H.~McWilliam, F.~Valentin, I.~M. Wallace, A.~Wilm, R.~Lopez, J.~D. Thompson, T.~J. Gibson, and D.~G. Higgins.
\newblock {Clustal W and Clustal X version 2.0}.
\newblock \emph{Bioinformatics}, 23\penalty0 (21):\penalty0 2947--2948, 11 2007.
\newblock ISSN 13674803.
\newblock \doi{10.1093/bioinformatics/btm404}.

\bibitem[Cock et~al.(2009)Cock, Antao, Chang, Chapman, Cox, Dalke, Friedberg, Hamelryck, Kauff, Wilczynski, and De~Hoon]{Cock2009Biopython:Bioinformatics}
Peter~J.A. Cock, Tiago Antao, Jeffrey~T. Chang, Brad~A. Chapman, Cymon~J. Cox, Andrew Dalke, Iddo Friedberg, Thomas Hamelryck, Frank Kauff, Bartek Wilczynski, and Michiel~J.L. De~Hoon.
\newblock {Biopython: Freely available Python tools for computational molecular biology and bioinformatics}.
\newblock \emph{Bioinformatics}, 25\penalty0 (11):\penalty0 1422--1423, 6 2009.
\newblock ISSN 13674803.
\newblock \doi{10.1093/bioinformatics/btp163}.

\bibitem[Altschup et~al.(1990)Altschup, Gish, Miller, Myers, and Lipman]{Altschup1990BasicTool}
Stephen~F Altschup, Warren Gish, Webb Miller, Eugene~W Myers, and David~J Lipman.
\newblock {Basic Local Alignment Search Tool}.
\newblock \emph{J. Mol. Biol}, 215:\penalty0 403--410, 1990.

\bibitem[Tareen and Kinney(2019)]{Tareen2019Logomaker:Python}
Ammar Tareen and Justin~B Kinney.
\newblock {Logomaker: Beautiful Sequence Logos in Python}.
\newblock \emph{biorxiv}, 2019.
\newblock \doi{10.1101/635029}.
\newblock URL \url{https://doi.org/10.1101/635029}.

\bibitem[{Schr\"odinger, LLC}(2015)]{PyMOL}
{Schr\"odinger, LLC}.
\newblock The {PyMOL} molecular graphics system, version~1.8.
\newblock November 2015.

\bibitem[Dauparas et~al.(2019)Dauparas, Wang, Swartz, Koo, Nitzan, and Ovchinnikov]{Dauparas2019UnifiedSequences}
Justas Dauparas, Haobo Wang, Avi Swartz, Peter Koo, Mor Nitzan, and Sergey Ovchinnikov.
\newblock {Unified framework for modeling multivariate distributions in biological sequences}.
\newblock \emph{arXiv}, 2019.
\newblock URL \url{http://arxiv.org/abs/1906.02598}.

\bibitem[Yang et~al.(2022)Yang, Eleutherai, and Yeh]{Yang2022MaskedLearning}
Kevin~K Yang, Niccolò~Zanichelli Eleutherai, and Hugh Yeh.
\newblock {Masked inverse folding with sequence transfer for protein representation learning}.
\newblock \emph{bioRxiv}, 2022.
\newblock \doi{10.1101/2022.05.25.493516}.
\newblock URL \url{https://www.biorxiv.org/content/early/2022/05/28/2022.05.25.493516}.

\bibitem[Koenig et~al.(2017)Koenig, Lee, Walters, Janakiraman, Stinson, Patapoff, and Fuh]{Koenig2017MutationalBinding}
Patrick Koenig, Chingwei~V. Lee, Benjamin~T. Walters, Vasantharajan Janakiraman, Jeremy Stinson, Thomas~W. Patapoff, and Germaine Fuh.
\newblock {Mutational landscape of antibody variable domains reveals a switch modulating the interdomain conformational dynamics and antigen binding}.
\newblock \emph{Proceedings of the National Academy of Sciences of the United States of America}, 114\penalty0 (4):\penalty0 E486--E495, 2017.
\newblock ISSN 10916490.
\newblock \doi{10.1073/pnas.1613231114}.

\bibitem[Warszawski et~al.(2019)Warszawski, Katz, Lipsh, Khmelnitsky, Nissan, Javitt, Dym, Unger, Knop, Albeck, Diskin, Fass, Sharon, and Fleishman]{Warszawski2019OptimizingInterfaces}
Shira Warszawski, Aliza~Borenstein Katz, Rosalie Lipsh, Lev Khmelnitsky, Gili~Ben Nissan, Gabriel Javitt, Orly Dym, Tamar Unger, Orli Knop, Shira Albeck, Ron Diskin, Deborah Fass, Michal Sharon, and Sarel~J. Fleishman.
\newblock {Optimizing antibody affinity and stability by the automated design of the variable light-heavy chain interfaces}.
\newblock \emph{PLoS Computational Biology}, 15\penalty0 (8):\penalty0 1--24, 2019.
\newblock ISSN 15537358.
\newblock \doi{10.1371/journal.pcbi.1007207}.

\bibitem[Hie et~al.(2022)Hie, Xu, Shanker, Bruun, Weidenbacher, Tang, and Kim]{Hie2022EfficientAlone}
Brian~L. Hie, Duo Xu, Varun~R. Shanker, Theodora~U.J. Bruun, Payton~A. Weidenbacher, Shaogeng Tang, and Peter~S. Kim.
\newblock {Efficient evolution of human antibodies from general protein language models and sequence information alone}.
\newblock \emph{bioRxiv}, page 2022.04.10.487811, 2022.
\newblock URL \url{https://www.biorxiv.org/content/10.1101/2022.04.10.487811v1%0Ahttps://www.biorxiv.org/content/10.1101/2022.04.10.487811v1.abstract}.

\end{thebibliography}

\newpage
\tableofcontents

\newpage
\appendix
\renewcommand\thefigure{\thesection.\arabic{figure}}
\renewcommand\thetable{\thesection.\arabic{table}}

\section{Appendix}

\setcounter{figure}{0}
\setcounter{table}{0}

\subsection{Method}
\subsubsection{Dataset}
\label{subsubsection:dataset_split}
\begin{figure}[h]
  \centering
  \includegraphics[width=0.7\linewidth]{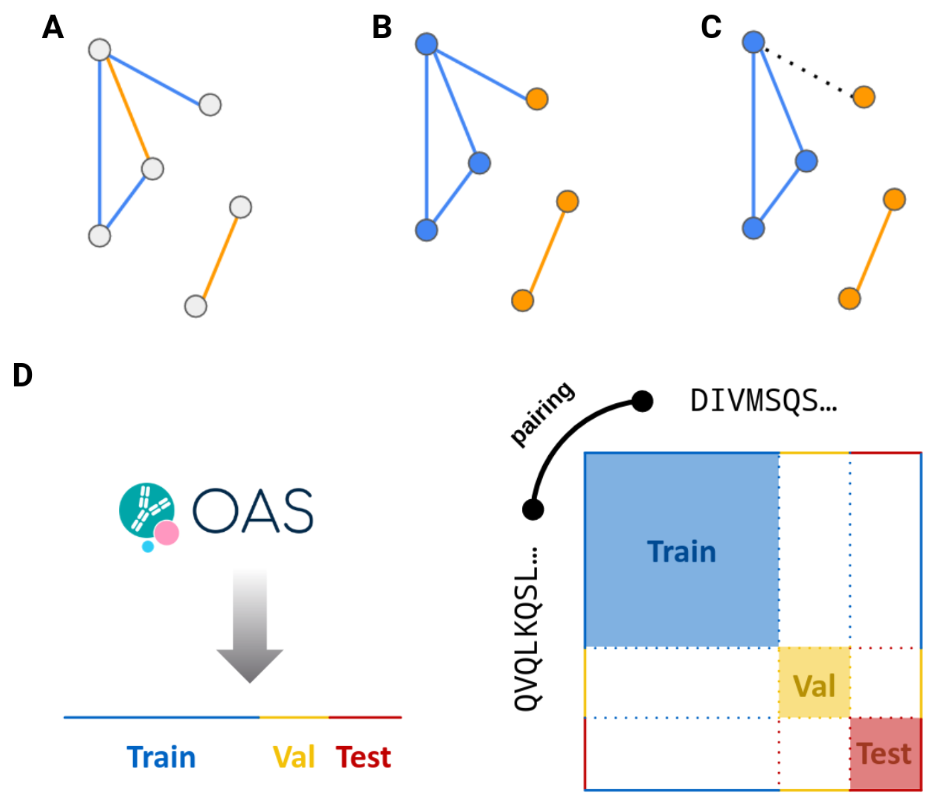}
  \caption{Splitting for protein-protein interaction dataset. Each sequence and pairing is represented by a node and an edge respectively, colorized by train (blue) and test (orange) partitions. (A) Interaction split. Nodes are not partitioned and are therefore colorless. (B) Inclusive node split. (C) Exclusive node split. Edges between train and test nodes are dropped (dotted line). (D) Exclusive node splitting in detail. All non-redundant sequences in paired OAS database are first split into train, validation, and test partitions. Only pairings within each partition are included in the final dataset, i.e. all cross-pairings are dropped.}
  \label{fig:graph_split}
\end{figure}

A visualization of dataset splitting strategy is given by Figure \ref{fig:graph_split}.

\subsection{Pairing Perplexity Reflects Preferences in Chain Pairing}
\label{appendix:pairing}

To demonstrate that our model understands the context of antibody pairing, we evaluate the model based on the perplexity of the sequence pairs. Using the human LM T5 in English-to-German translation as an analogy, feeding an English sentence to the encoder and its German counterpart to the decoder should in general yield a lower perplexity than feeding both encoder and decoder with English sentences. The idea is to probe the model's capability to understand that a German sentence should be generated from an English input in a generative model, instead of assessing the model in a traditional sentence-pair classification task.

Without publicly available antibody mispairing dataset, we test our model on two simple mispairing scenarios, i.e. chain-type mispairing and species mispairing. For chain-type mispairing, we synthesize \textit{correct} heavy-light pairing and \textit{mispaired} heavy-heavy/light-light pairing for each translation in test set, with the assumption that only heavy-light-chain pairings are permitted. A similar approach is used for species mispairing by assuming cross-species chain pairing is impermissible. Note that, given the promiscuous nature of antibody chain pairing, a heavy chain sequence can pair with multiple light chain chain sequences. Therefore, randomly paired heavy and light chain can still be a valid pairing and cannot serve as a negative control in comparison to observed pairing by contrasting their respective perplexity.

\begin{figure}[h]
  \centering
  \includegraphics[width=0.9\linewidth]{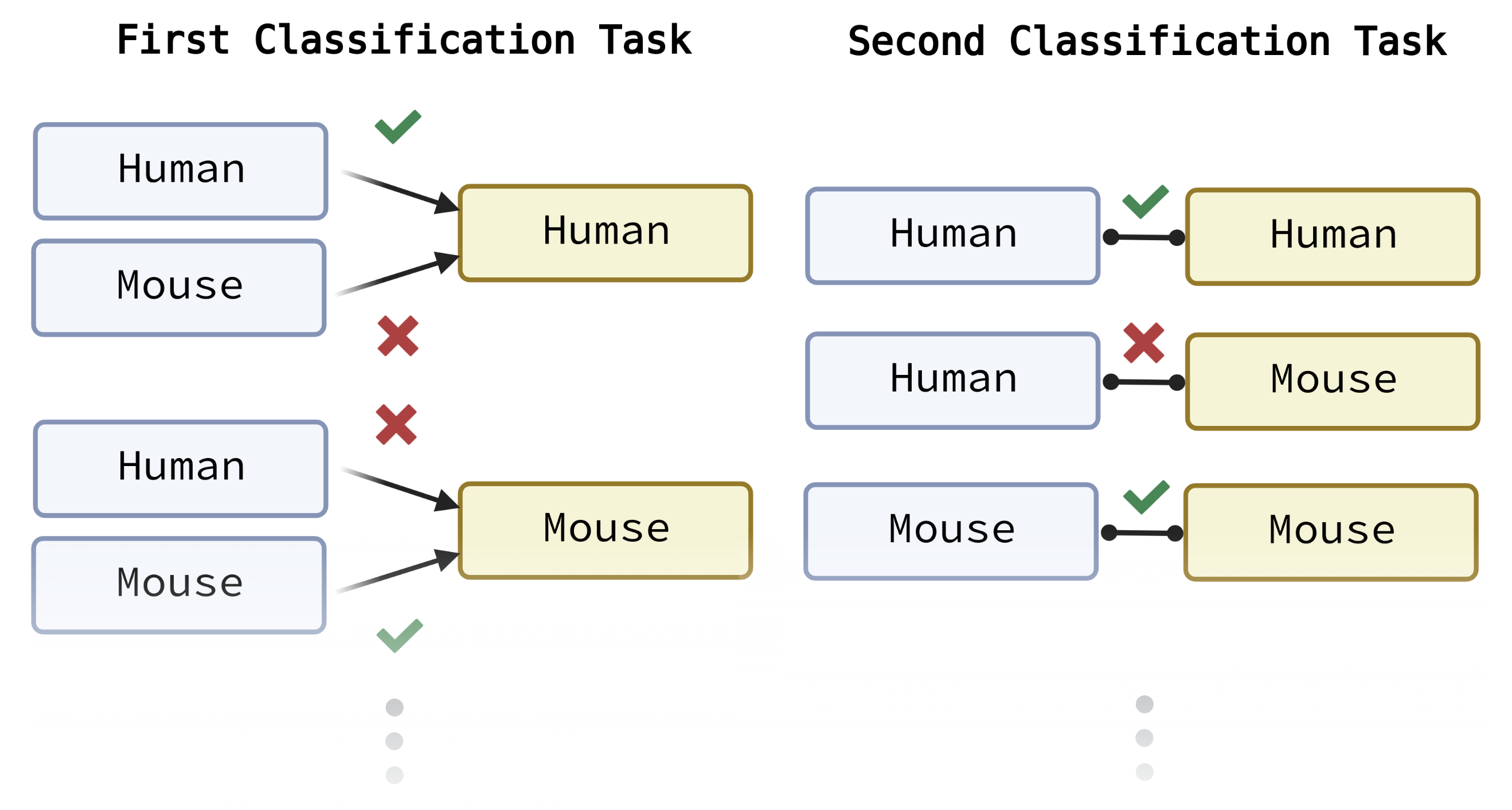}
  \caption{Schematics of two classification tasks considered for species mispairing. (Left) In the first classification task, the aim is to identify the \textit{correct} and \textit{mispaired} sequences sharing the same target. (Right) In the second classification task, the aim is to predict the likelihood of the pairing as a bidirectional translation. The tasks for chain-type mispairing are similar. No chain type nor species annotation is used in our prediction.}
  \label{fig:classification_tasks}
\end{figure}

We propose two classification tasks (Figure \ref{fig:classification_tasks}) to assess our model's perplexity. The first task considers two input sequences sharing the same target sequence and only one pairing is correct. Out of the two pairings, we assign the pairing with lower perplexity as \textit{correct} and the other one as \textit{mispaired}. Based on this assignment, we identify above 90\% of the \textit{correct} pairings from chain-type mispairing and close to 80\% from species mispairing. The baseline of random assignment results in 50\% accuracy. No classification model is trained.

In our second task, we consider a dataset by mixing and shuffling the \textit{correct} and \textit{mispaired} samples from the first task and classify whether the pairing is \textit{correct} given two antibody sequences alone. Informed only by our language model’s perplexity, a logistic regression significantly outperforms the baseline of random assignment. The classifier is trained on the average perplexity of forward- and back-translations on validation set. All performance metrics are evaluated on test set. The weaker classification performance might be attributed to the loss of pairing preferences between gene loci and families in the creation of mispairing dataset (Subsection \ref{appendix:pairing}).

\begin{table}[h]
  \centering
  \begin{minipage}{0.9\linewidth}
    \centering
      \begin{tabu}{c c c}
        \multicolumn{3}{c}{\textbf{First Classification Task}} \\
        \hline
        Mispairing type & Target chain & Accuracy \\
        \hline
        \multirow{2}{*}{Chain type} & Light & 0.92 \\
        \cline{2-3} & Heavy & 0.91 \\
        \hline
        \multirow{2}*{Species} & Light & 0.80 \\
        \cline{2-3} & Heavy & 0.79 \\
        \hline
    \end{tabu}
    \vspace{2mm}
    \end{minipage}
    \begin{minipage}{0.9\linewidth}
    \centering
      \begin{tabu}{c c c}
      \multicolumn{3}{c}{\textbf{Second Classification Task}} \\
      \hline
      Mispairing type & Accuracy & AUROC \\
      \hline
      Chain type & 0.54 & 0.70 \\
      Species & 0.57 & 0.60 \\
      \hline
    \end{tabu}
    \vspace{2mm}
    \end{minipage}
  \caption{Performance on first and second classification task on model perplexity alone. (Up) In the first classification task, mispairing assignment is based on the rank of perplexity without any parameterizable model. (Bottom) In the second classification task, logistic regression is trained on the bidirectional average of translation perplexity in validation set, and evaluated on test set. Random assignment results in an accuracy of 0.5 in the first task, and an additional AUROC of 0.5 in the second task.}
\end{table}

To further elaborate on the methodology, we generate synthetic mispairings to test our model’s capability of learning chain pairing. The generation protocol for chain-type mispairing is as follows (algorithm \ref{algorithm:chaintype_mispairing}). The generation protocol for species mispairing is similar (algorithm \ref{algorithm:species_mispairing}).

\begin{algorithm}
\caption{Chain-type mispairing dataset generation}
\label{algorithm:chaintype_mispairing}
\begin{algorithmic}[1]
\State Inputs: paired test dataset $D$
\State Outputs: chain-type mispairing dataset $D'$
\State initialize $H$, $L$ and $D'$ as $\emptyset$
\For{($u$, $v$) in $D$}
   \For{$s$ in ($u$, $v$)}
      \If{chaintype($s$) = heavy}
        \State $H$.add($s$)
      \ElsIf{chaintype($s$) = light}
        \State $L$.add($s$)
      \EndIf
   \EndFor
\EndFor

\For{($u$, $v$) in $D$}
  \If{chaintype($u$) = chaintype($v$)}
    \For{$s$ in ($u$, $v$)}
      \If{chaintype($s$) = heavy}
        \State $s' \gets random\ element\ in\ L$
      \ElsIf{chaintype($s$) = light}
        \State $s' \gets random\ element\ in\ H$
      \EndIf
      \State $D'$.add(($s$, $s'$))
    \EndFor
  \EndIf
\EndFor
\State \Return $D'$
\end{algorithmic}
\end{algorithm}

\begin{algorithm}
\caption{Species mispairing dataset generation}
\label{algorithm:species_mispairing}
\begin{algorithmic}[1]
\State Inputs: paired test dataset $D$
\State Outputs: species mispairing dataset $D'$
\State initialize $H$, $M$ and $D'$ as $\emptyset$
\For{($u$, $v$) in $D$}
   \For{$s$ in ($u$, $v$)}
      \If{species($s$) = human}
        \State $H$.add($s$)
      \ElsIf{species($s$) = mouse}
        \State $M$.add($s$)
      \EndIf
   \EndFor
\EndFor

\For{($u$, $v$) in $D$}
  \If{species($u$) = species($v$)}
    \For{$s$ in ($u$, $v$)}
      \If{species($s$) = human}
        \State $s' \gets random\ element\ in\ M$
      \ElsIf{species($s$) = mouse}
        \State $s' \gets random\ element\ in\ H$
      \EndIf
      \State $D'$.add(($s$, $s'$))
    \EndFor
  \EndIf
\EndFor
\State \Return $D'$
\end{algorithmic}
\end{algorithm}

\begin{figure}[H]
  \centering
  \includegraphics[width=\linewidth]{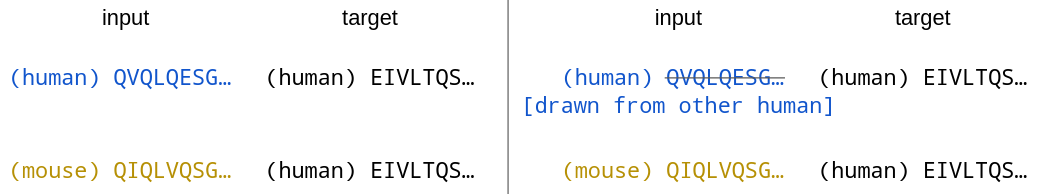}
  \caption{Schematics of preparation of correct and mispaired sequences in species mispairing. The input sequence for correct pairing is in blue and that for mispairing is in yellow. (Left) Single-generation scheme: comparison between observed correct pairing and synthetic mispairing. (Right) Double-generation scheme: comparison between synthetic \textit{correct} pairing and synthetic mispairing.}
  \label{fig:mispairing_scheme}
\end{figure}

We have considered two possible schemes for preparing correct pairings (Figure \ref{fig:mispairing_scheme}), i.e. single-generation and double-generation. In single-generation, we keep the observed pairing from test set as the correct pairing. While it ensures that the correct pairing is experimentally validated, the comparison between an observed correct pairing and a synthetic mispairing creates a bias in perplexity.

As such, we introduce double-generation where both pairings are generated and label the synthetically \textit{correct} pairing in italic. Despite the lack of direct experiment validation, the comparison between correct and mispaired pairings is unbiased, is more challenging than single-generation, and provides some insights into whether our model learns antibody chain pairing. As indicated in Table \ref{tab:single-generation1} and \ref{tab:single-generation2}, the conclusion remains the same when switched from single-generation to double-generation.

\begin{table}[h]
\centering
{\tabulinesep=1.1mm
\begin{tabu}{c c c}
  \hline
  Mispairing type & Target chain & Accuracy \\
  \hline
  \multirow{2}{*}{Chain type} & Light & 0.99 \\
  \cline{2-3} & Heavy & 0.96 \\
  \hline
  \multirow{2}*{Species} & Light & 0.97 \\
  \cline{2-3} & Heavy & 0.96 \\
  \hline
\end{tabu}}
\vspace{2mm}
\caption{First classification task assignment accuracy by the perplexity rank between correct and \textit{mispaired} antibody sequences in single-generation scheme.}
\label{tab:single-generation1}
\end{table}

\begin{table}[h]
\centering
{\tabulinesep=1.1mm
\begin{tabu}{c c c}
  \hline
  Mispairing type & Accuracy & AUROC \\
  \hline
  Chain-type & 0.54 & 0.72 \\
  Species & 0.60 & 0.70 \\
  \hline
\end{tabu}}
\vspace{2mm}
\caption{Second classification task performance in single-generation scheme}
\label{tab:single-generation2}
\end{table}

\subsection{Conditional Generation Recovers Pairing Sequences}
\label{appendix:representation}
\begin{figure}[h]
  \centering
  \includegraphics[width=0.7\linewidth]{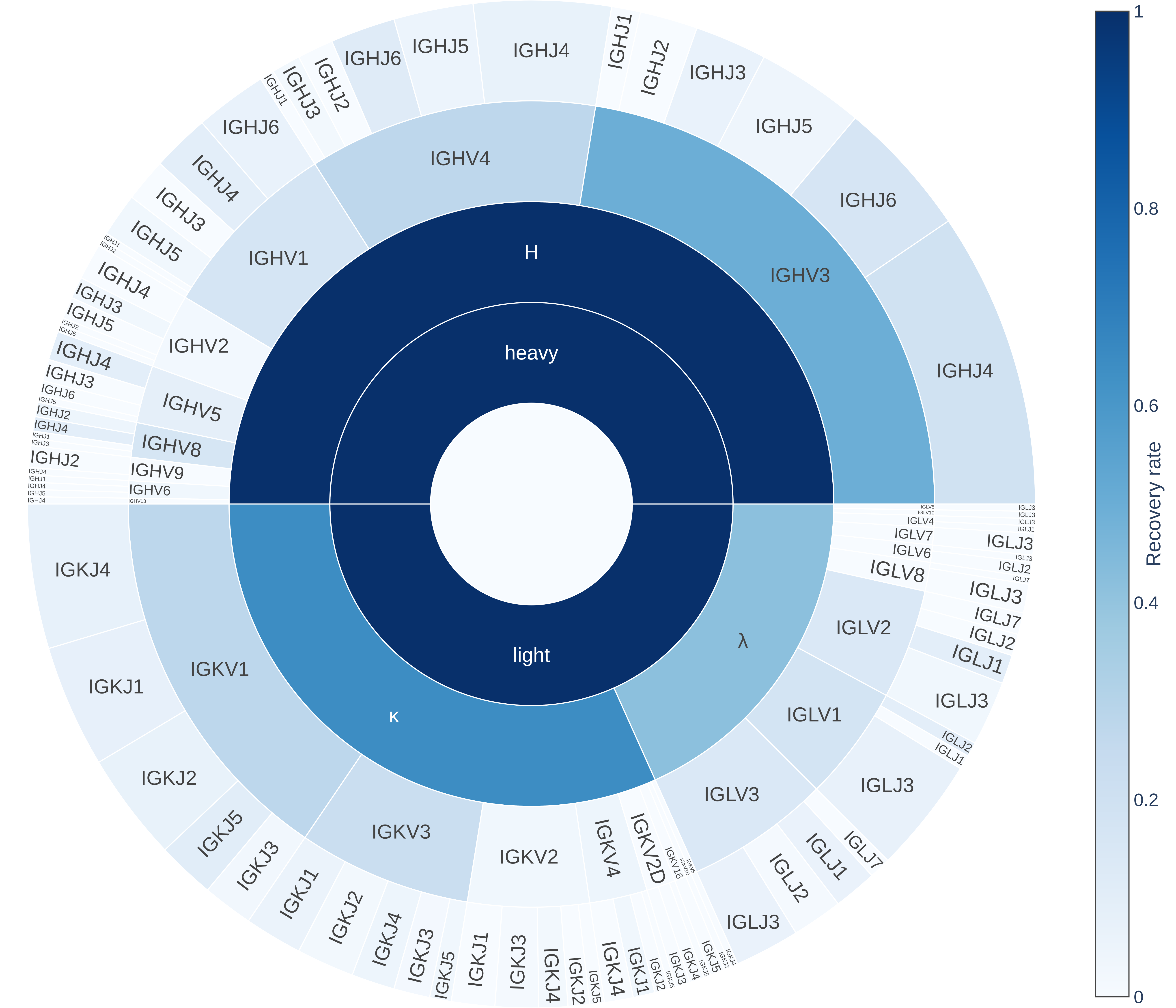}
  \caption{Recovery rate of target chain type, gene loci, and gene families in sequence generation. Performance is represented in a hierarchical order, where parent classes are centered while children categories are on the periphery. On each rim, the arc lengths of categories are proportional to their populations in test set. Dark blue represents perfect recovery whereas white color implies low recovery rate.}
  \label{fig:generation_sunbursts}
\end{figure}

\begin{figure}[H]
  \centering
  \includegraphics[width=0.75\linewidth]{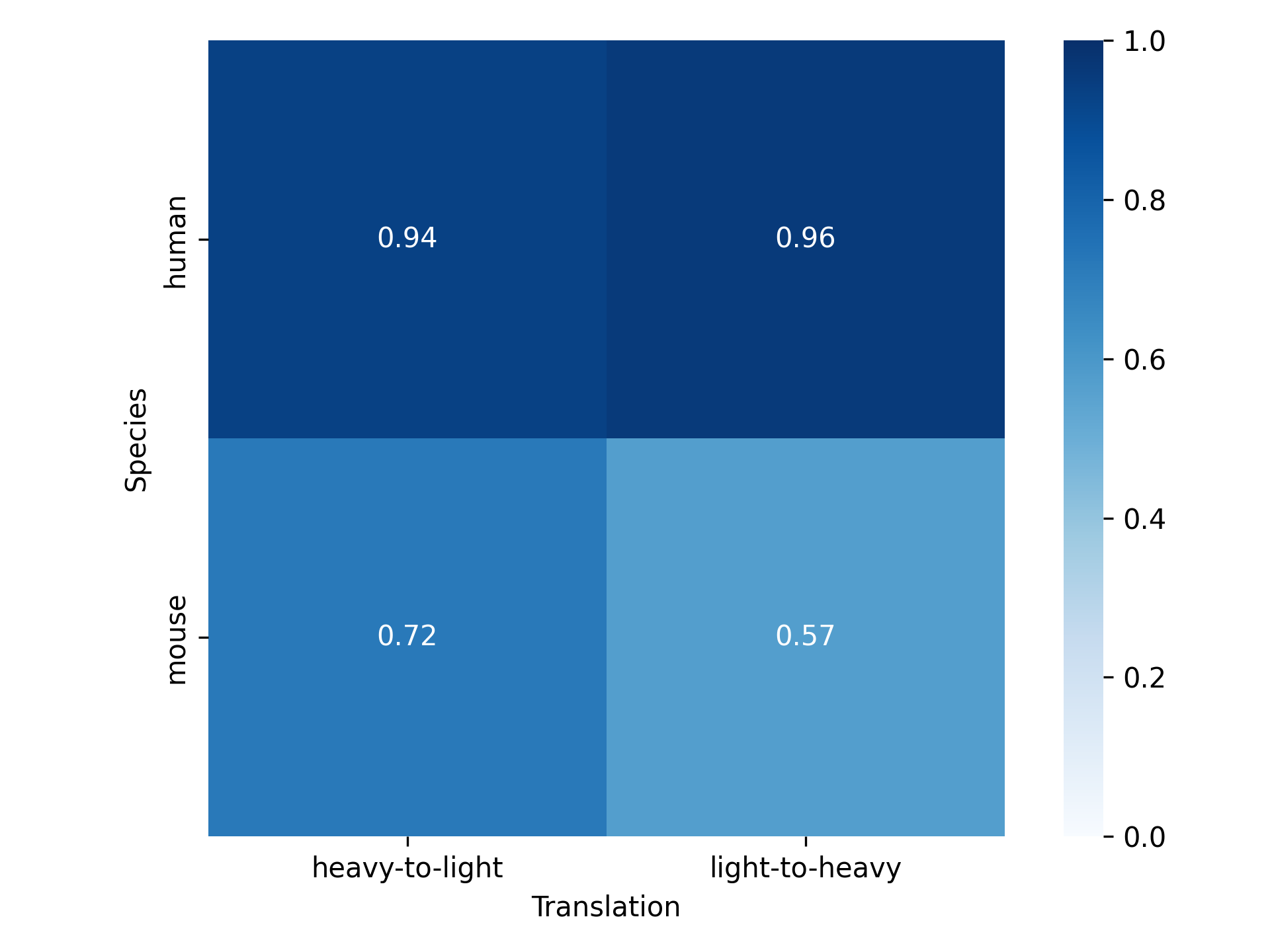}
  \caption{Recovery rate on species by original species and translation direction.}
  \label{fig:recovery_species}
\end{figure}

\begin{figure}[H]
  \centering
  \includegraphics[width=0.75\linewidth]{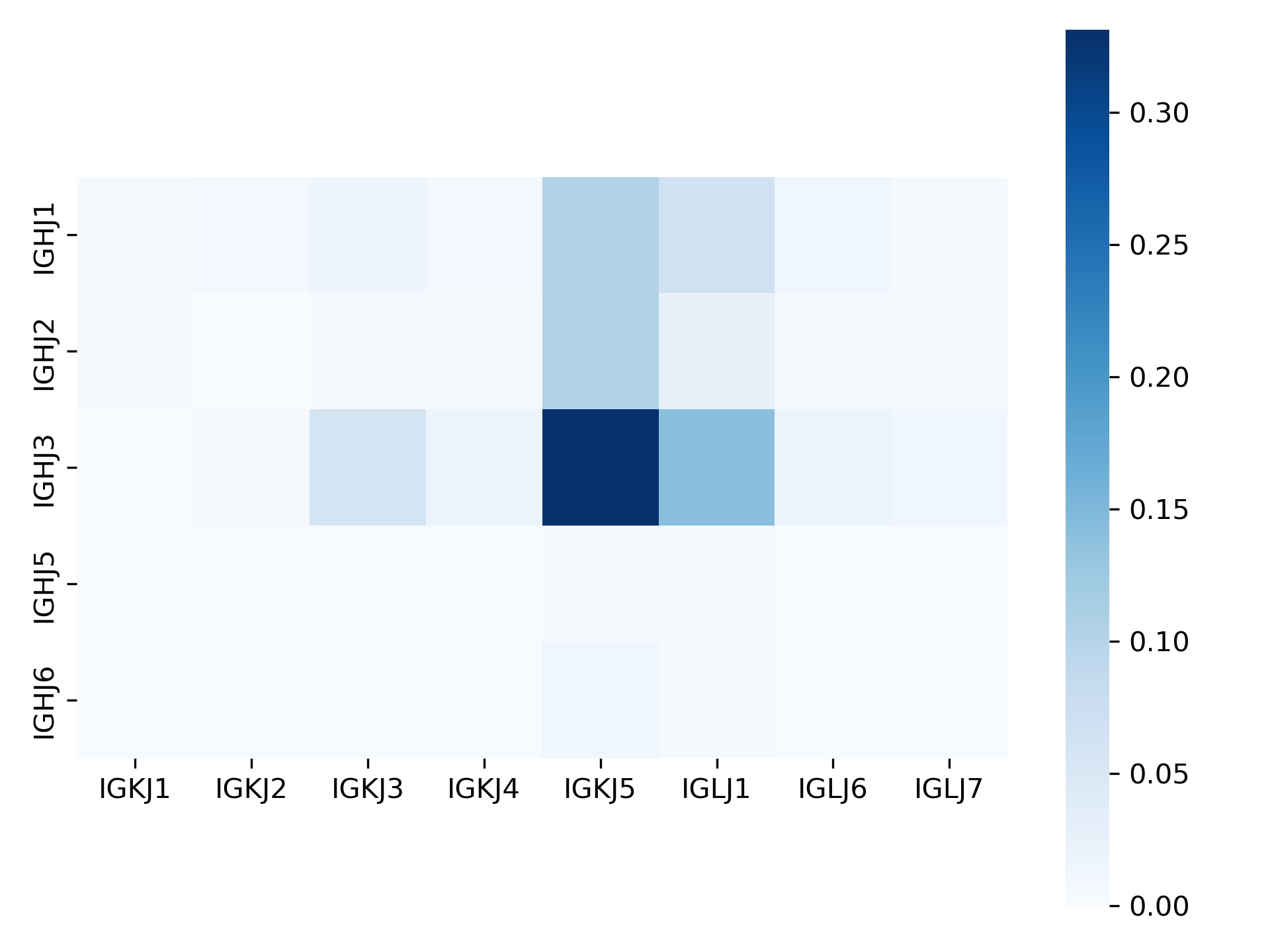}
  \caption{Co-occurrence of J families in heavy and light chains colorized by relative frequency. Frequency is normalized by the total number of observed co-occurrence.}
  \label{fig:jheavy_jlight}
\end{figure}

In order to evaluate our model’s sequence-to-sequence generative performance, we test whether our model can recover the observed pairing in test set. Figure \ref{fig:generation_sunbursts} illustrates the recovery rate at progressively fine levels of resolution on human antibodies. A target sequence is considered to be recovered if the generated sequence shares the same chain type, gene loci, V gene family, or the combination of V and J gene families. For chain types, our model always generates heavy chains from light chain inputs, and likewise for light chain generation. For gene loci on light chain, $\lambda$ and $\kappa$ loci are recovered at 48\% and 56\% of the time. As we approach finer resolutions, the recovery rate drops in V families and their combination with J families. This is consistent with the observation that antibody chain pairing is often degenerate. For instance, the heavy chain sequences from IGHV1 gene family are observed to pair with multiple families in both $\lambda$ and $\kappa$ loci (Figure \ref{fig:jheavy_jlight}). This sets an upper bound on the recovery rate in antibody heavy and light chain pairing. A similar analysis has also been performed on the recovery of species (Figure \ref{fig:recovery_species}) and the exact figures of recovery rate can depend on the generative parameters, which are listed in Subsection \ref{appendix:representation}.

\begin{figure}[h]
    \centering
    \includegraphics[width=0.8\linewidth]{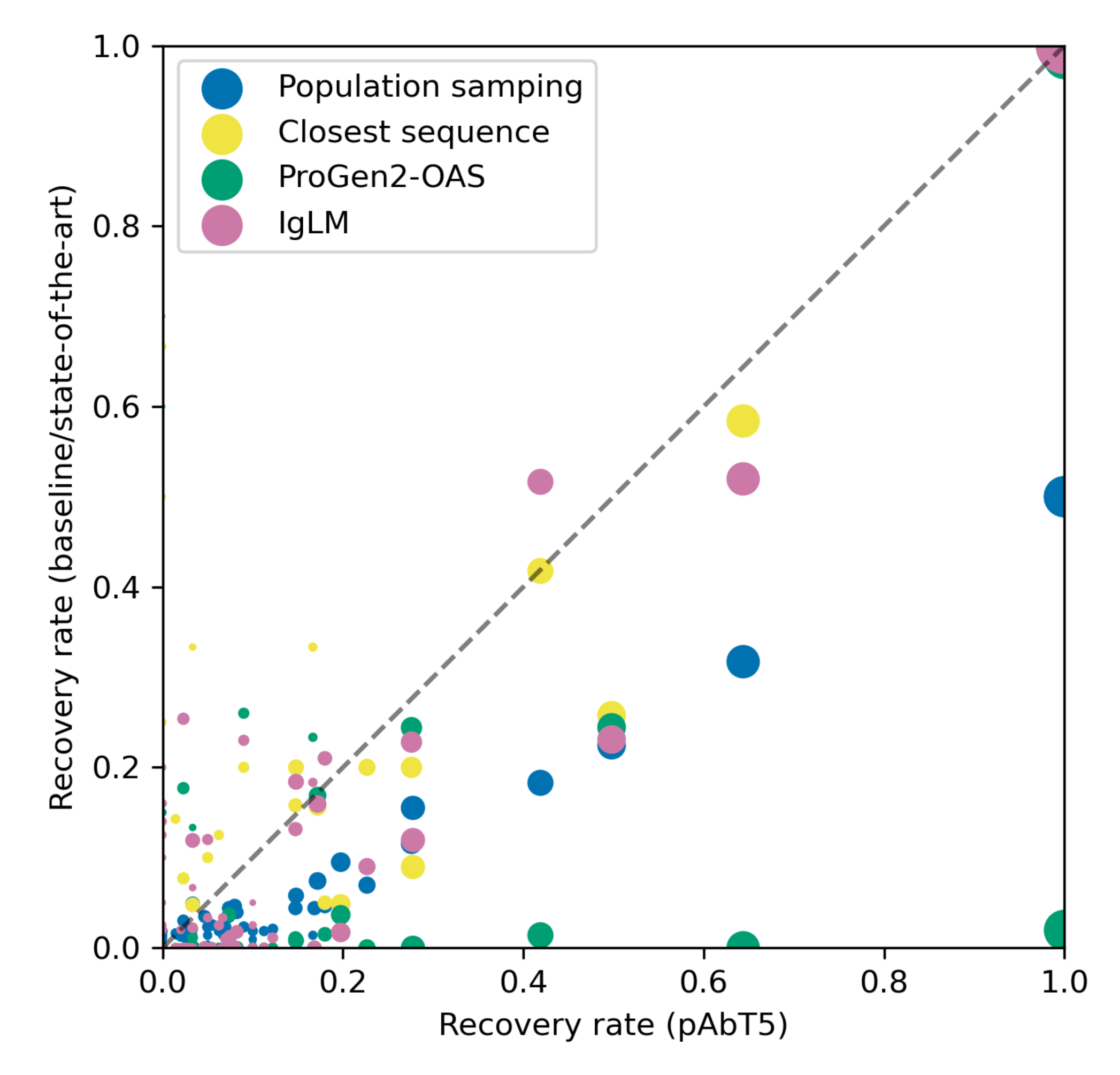}
    \caption{Percentage of generated sequences sharing the same chain type, gene loci, V gene families, and V-J gene families with target. x-axis is the recovery rate of pAbT5, and y-axis is the recovery rate of ProGen2-OAS \cite{Nijkamp2022ProGen2:Models}, IgLM \cite{Shuai2022GenerativeDesign}, picking a sequence randomly from the population, and picking the pairing partner of the closest sequences from the validation set. Each scatter point represents recovery at a resolution, and size of the scatter point is proportional to its respective population size. The full table is available in Supplementary Materials.}
    \label{fig:generation_baseline}
\end{figure}

We use ANARCI \cite{Dunbar2016ANARCI:Classification} for species, chain type, and gene family classification. Although OAS dataset indicates humans, mice, and rats as the source organisms, ANARCI identifies only the former two. For consistent comparison in both observed and generated antibody pairs, we opt for the definition in ANARCI in all evaluations, including t-SNE, mispairing, and generation assessment. We only report V and J families in heavy and light chains as D families are not supported by ANARCI. In all species-specific analyses, pairings are included only when ANARCI identifies both heavy and light chains from the same species.

We denote the encoder sequence as the input of the translation and decoder sequence as the target of the translation. We denote the encoder hidden state of the paired antibody in the translation order of input-to-target as the sequence embedding of the input sequence, or simply sequence embedding. For t-SNE visualization, we take the mean of the encoder hidden state over residues at the final layer.

In the generative process, sequences are generated at a temperature of 1, top p of 0.9 with 10 returned sequences, determined from a grid search of temperature and top p. Experiment on beam search results in low diversity and regions of repetitive motifs. All co-occurrences of gene families are collected from test set. For ProGen2-OAS \cite{Nijkamp2022ProGen2:Models}, we use the default generative parameters and do not provide the first few tokens to avoid hinting at the chain type and gene loci. We use default generative parameters in IgLM \cite{Shuai2022GenerativeDesign}. 

\begin{figure*}[h]
  \centering
  \includegraphics[width=0.7\textwidth]{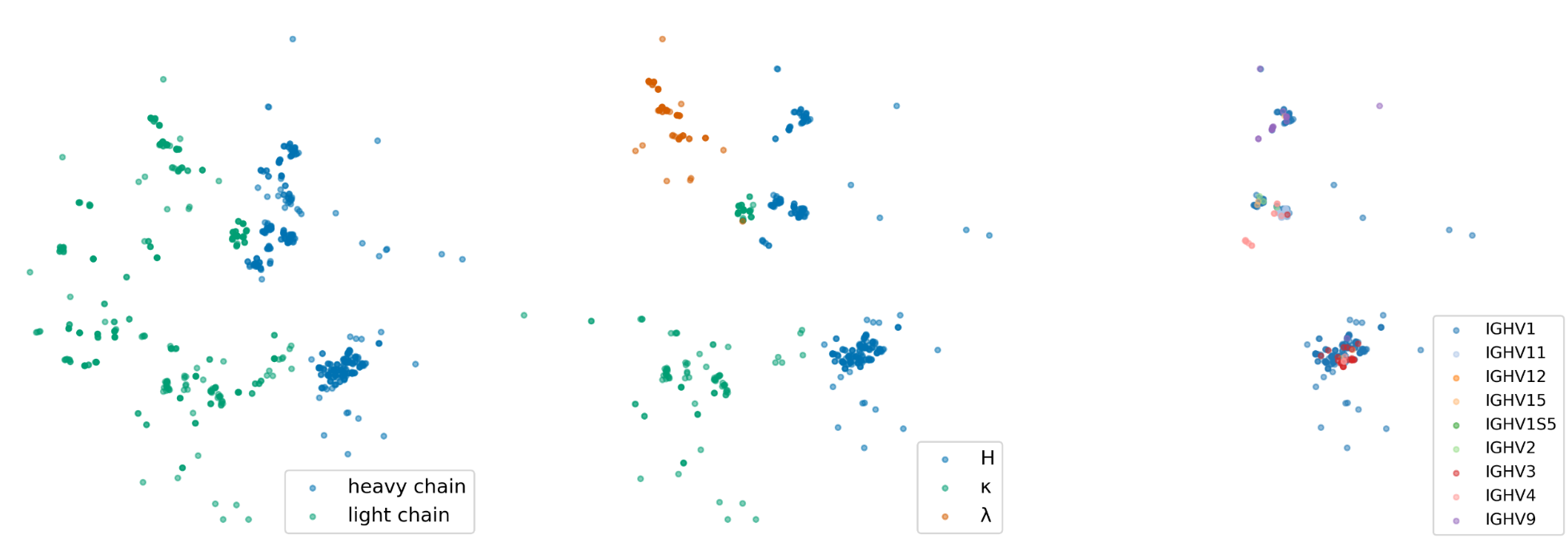}
  \caption{t-SNE plot of encoder hidden states of test set sequences in progressively fine categories (chain types, human gene loci, and human IGHV gene families).}
  \label{fig:hidden_state}
\end{figure*}

\begin{figure}[H]
  \centering
  \includegraphics[width=0.45\linewidth]{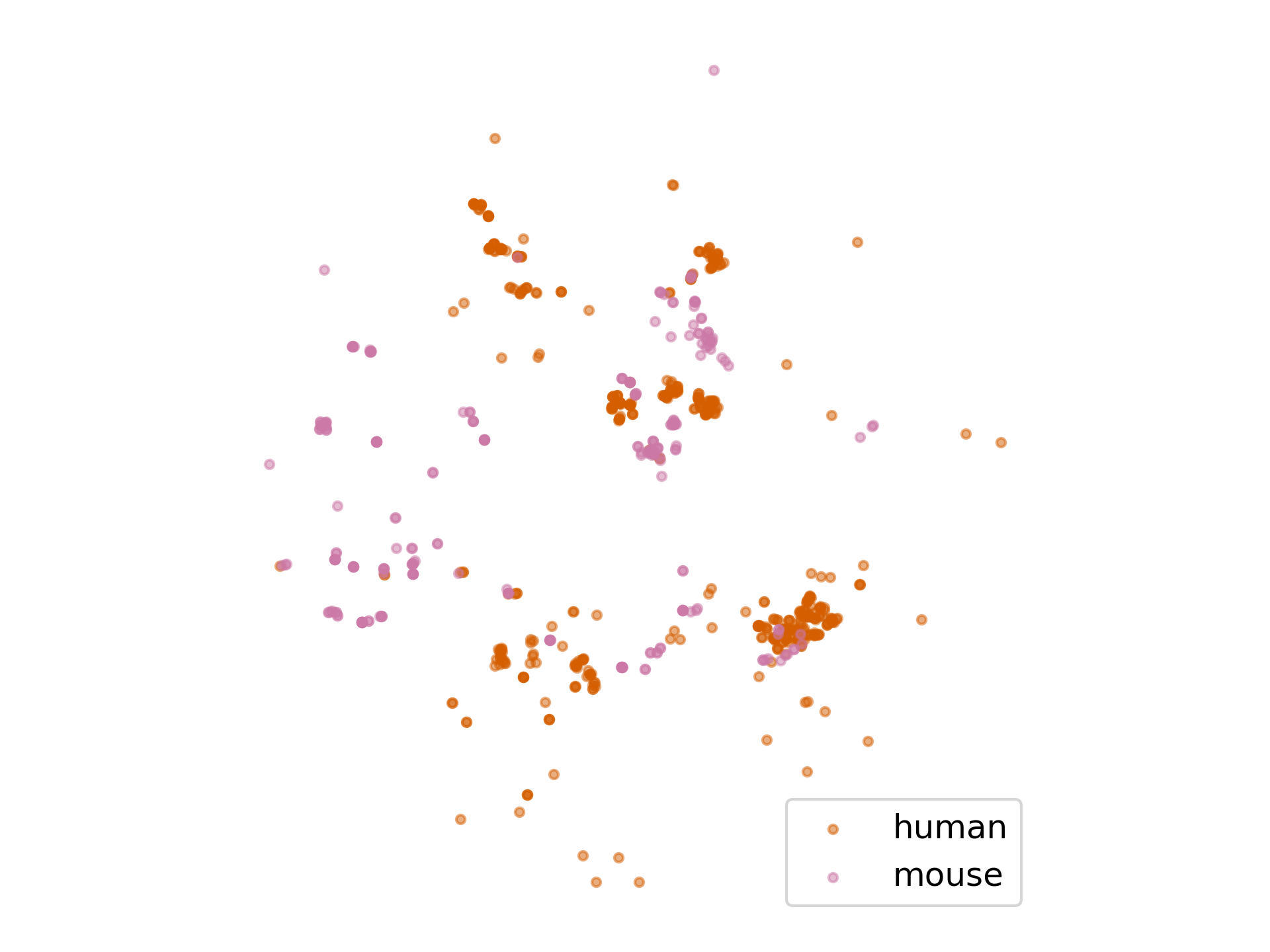}
  \caption{t-SNE plot of sequence embeddings colorized by ANARCI annotated species}
  \label{fig:tsne_species}
\end{figure}

\begin{figure}[H]
  \centering
  \includegraphics[width=\textwidth]{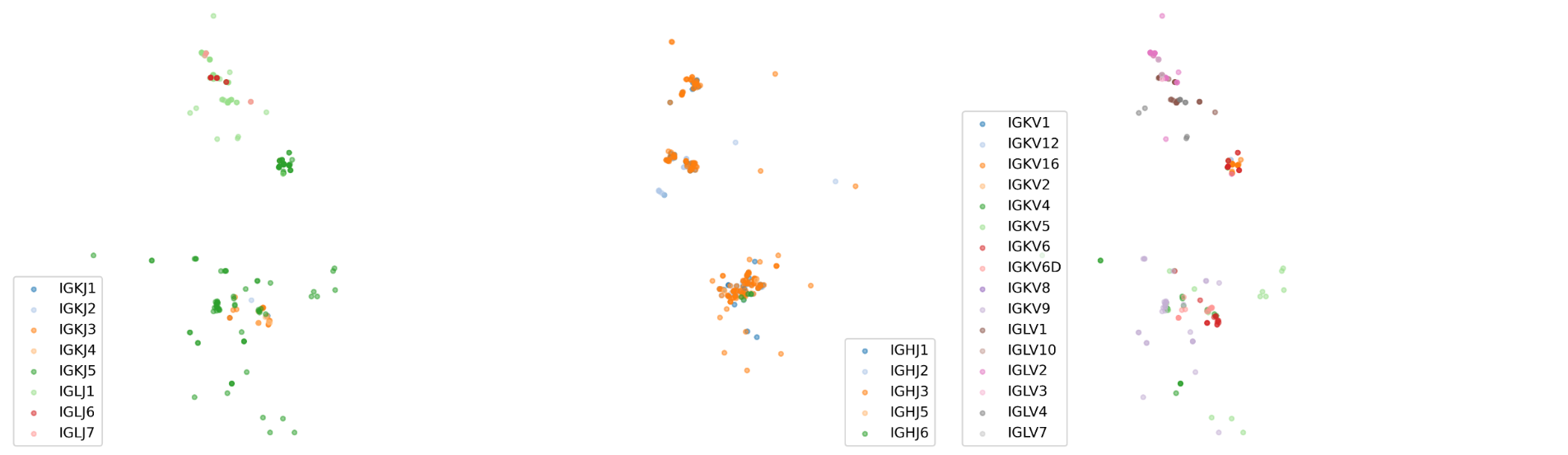}
  \caption{t-SNE plot of sequence embeddings colorized by ANARCI annotated gene families. (Left) Light chain J gene. (Middle) Heavy chain J gene. (Right) Light chain V gene.}
\end{figure}

\begin{figure}[H]
  \centering
  \includegraphics[width=0.75\linewidth]{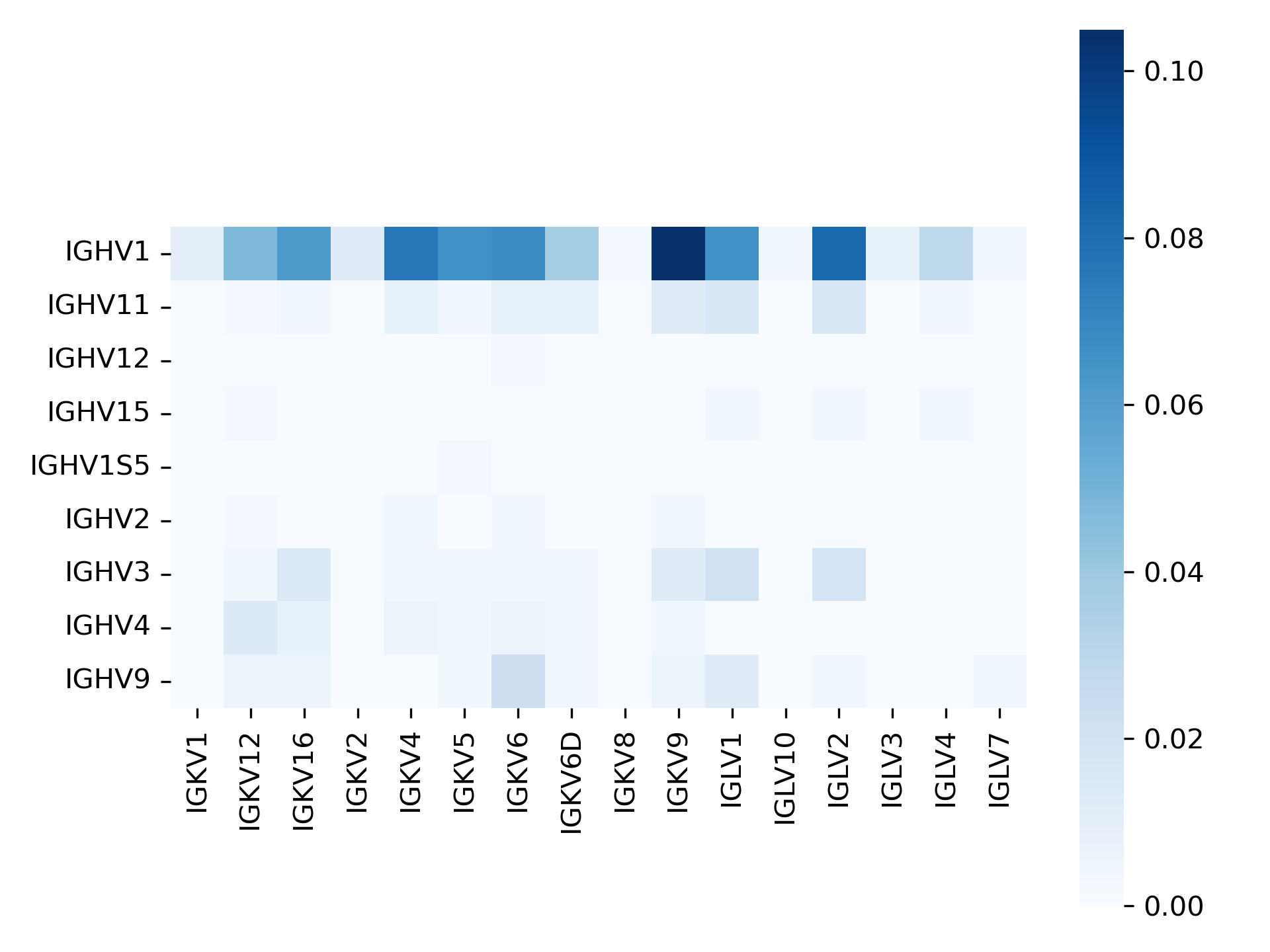}
  \caption{Co-occurrence of V families in heavy and light chains colorized by relative frequency. Frequency is normalized by the total number of observed co-occurrence.}
  \label{fig:vheavy_vlight}
  \end{figure}

\begin{figure}[H]
  \centering
  \includegraphics[width=0.75\linewidth]{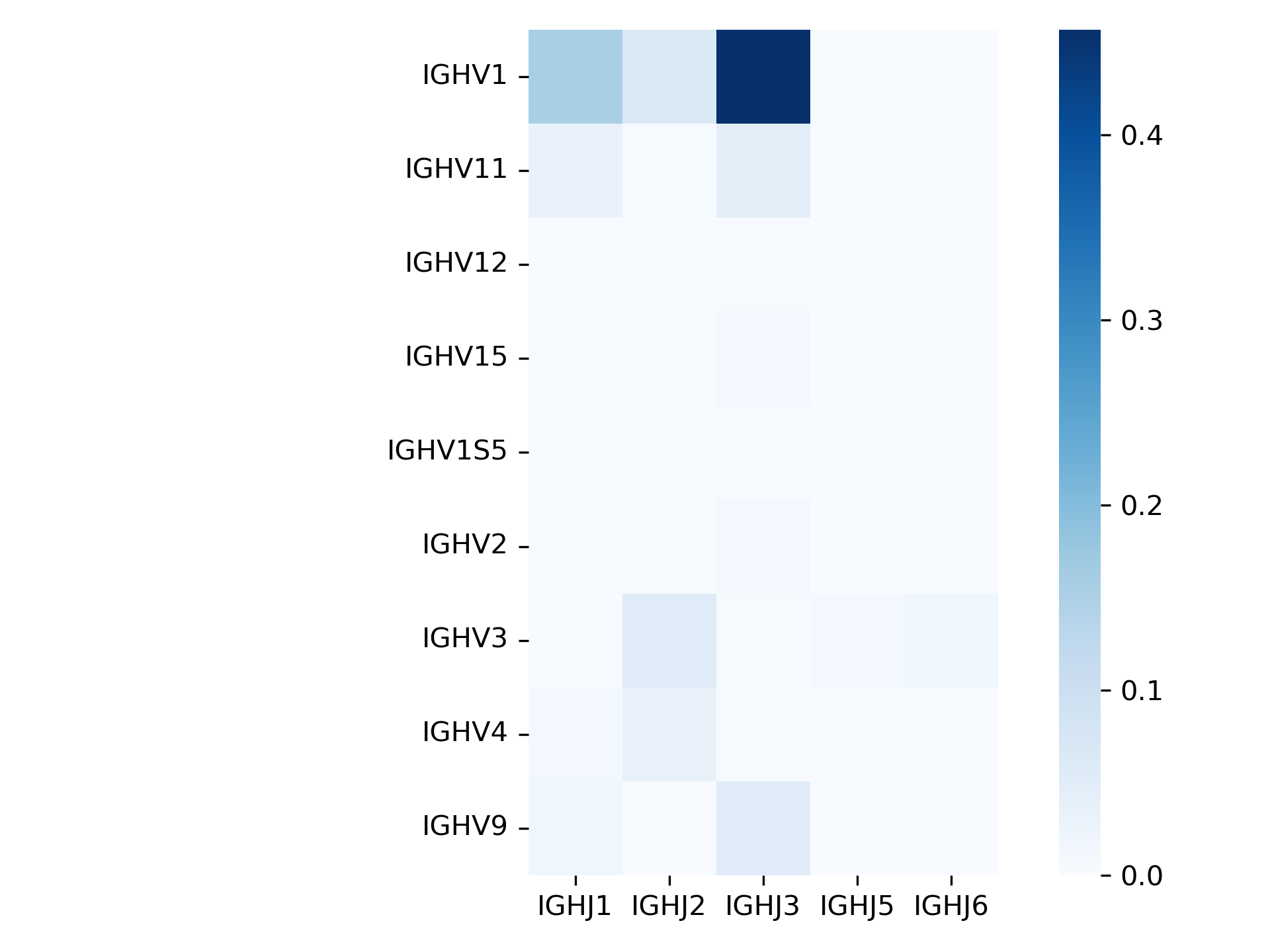}
  \caption{Co-occurrence of V and J families in heavy chain colorized by relative frequency. Frequency is normalized by the total number of observed co-occurrence.}
  \label{fig:vheavy_jheavy}
\end{figure}

\begin{figure}[H]
  \centering
  \includegraphics[width=0.75\linewidth]{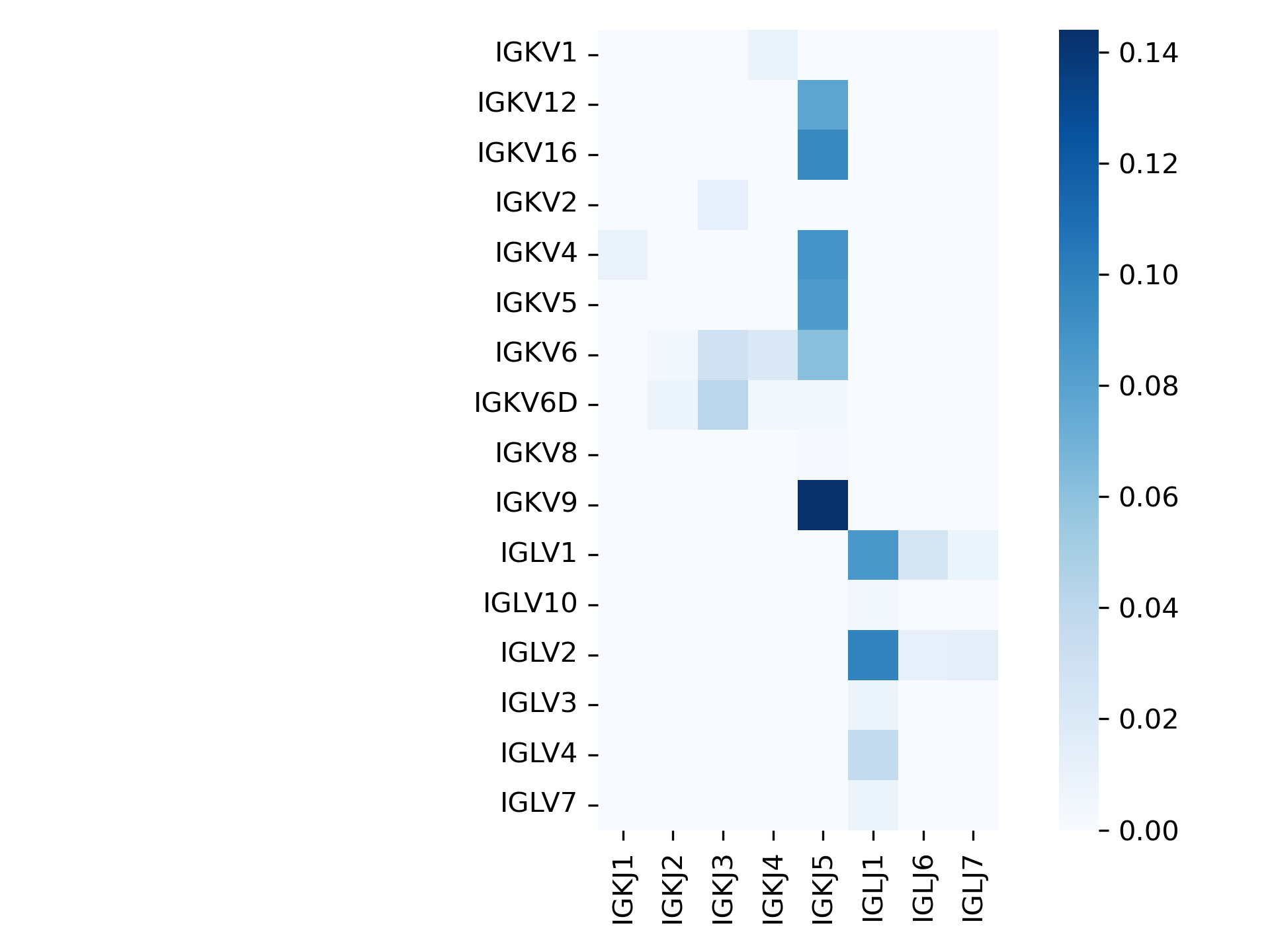}
  \caption{Co-occurrence of V and J families in light chain colorized by relative frequency. Frequency is normalized by the total number of observed co-occurrence.}
  \label{fig:vlight_jlight}
\end{figure}

\begin{figure}[H]
  \centering
  \includegraphics[width=0.75\linewidth]{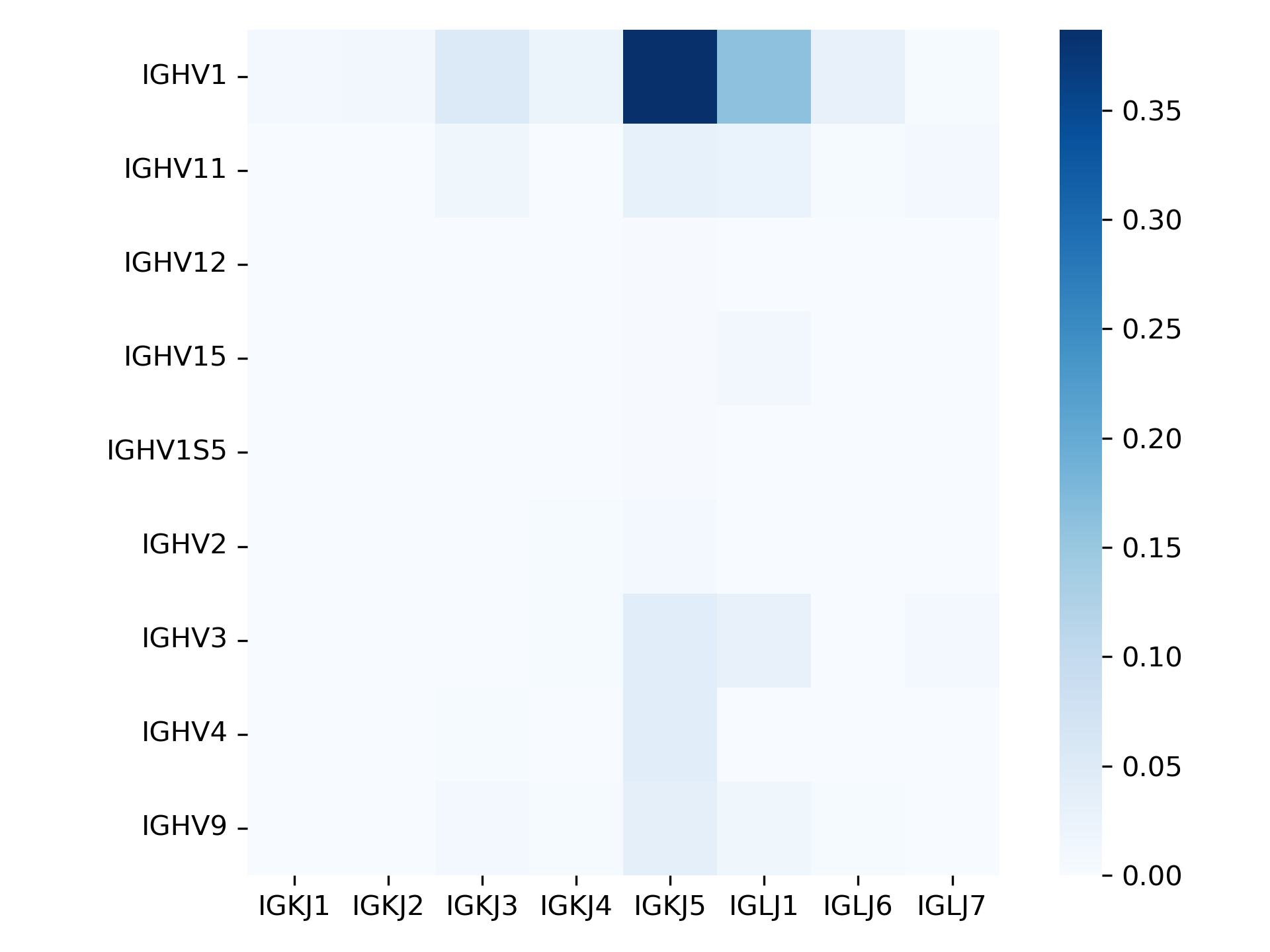}
  \caption{Co-occurrence of V families in heavy chain and J families in light chain colorized by relative frequency. Frequency is normalized by the total number of observed co-occurrence.}
  \label{fig:vheavy_jlight}
\end{figure}

\begin{figure}[H]
  \centering
  \includegraphics[width=0.75\linewidth]{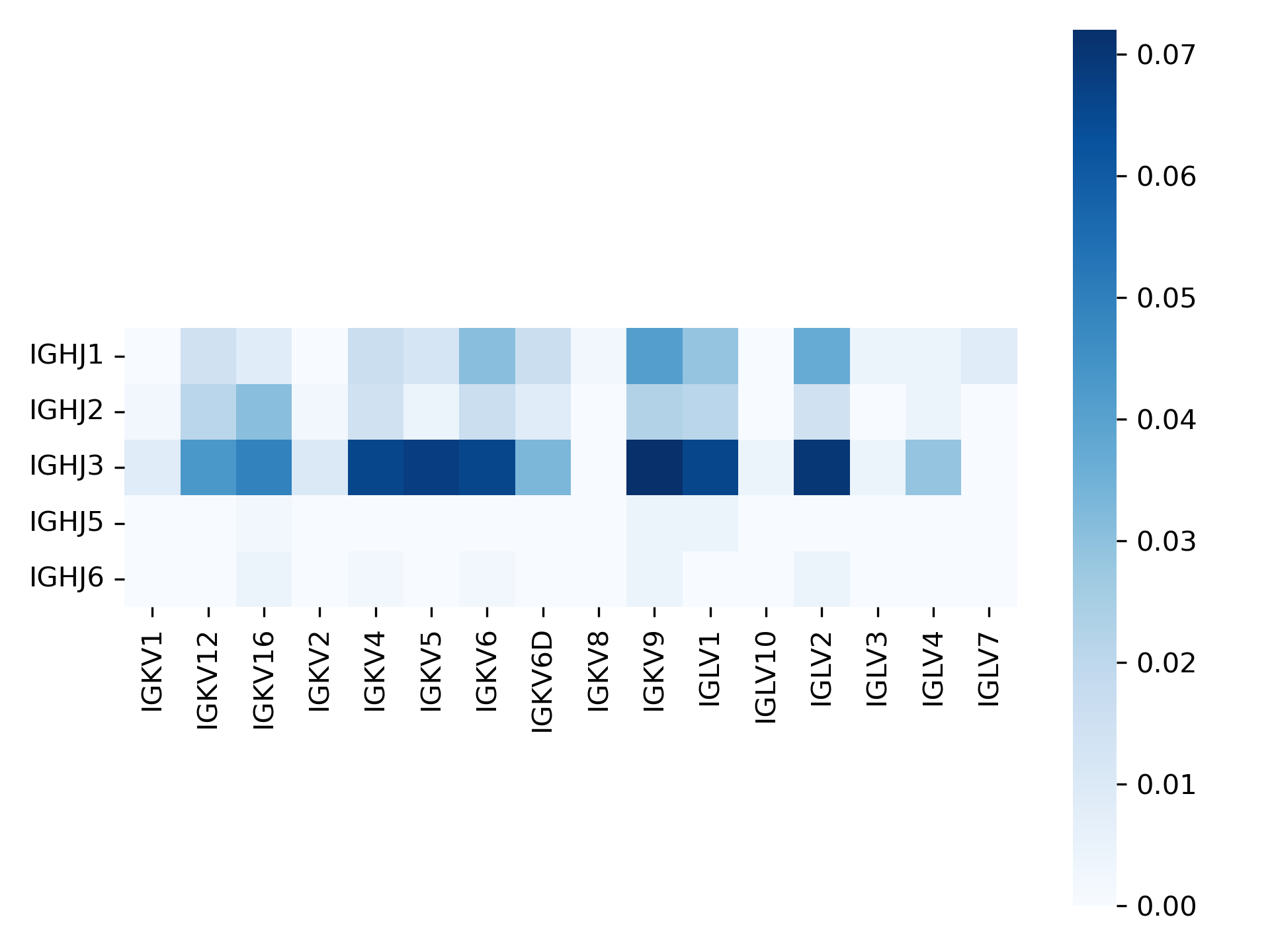}
  \caption{Co-occurrence of J families in heavy chain and J families in light chain colorized by relative frequency. Frequency is normalized by the total number of observed co-occurrence.}
  \label{fig:jheavy_vlight}
\end{figure}

\subsection{Sequence Generation Aligns with Conserved and Variable Domains in Antibodies}
\label{appendix:structures}
\begin{figure}[h]
  \centering
  \includegraphics[width=\linewidth]{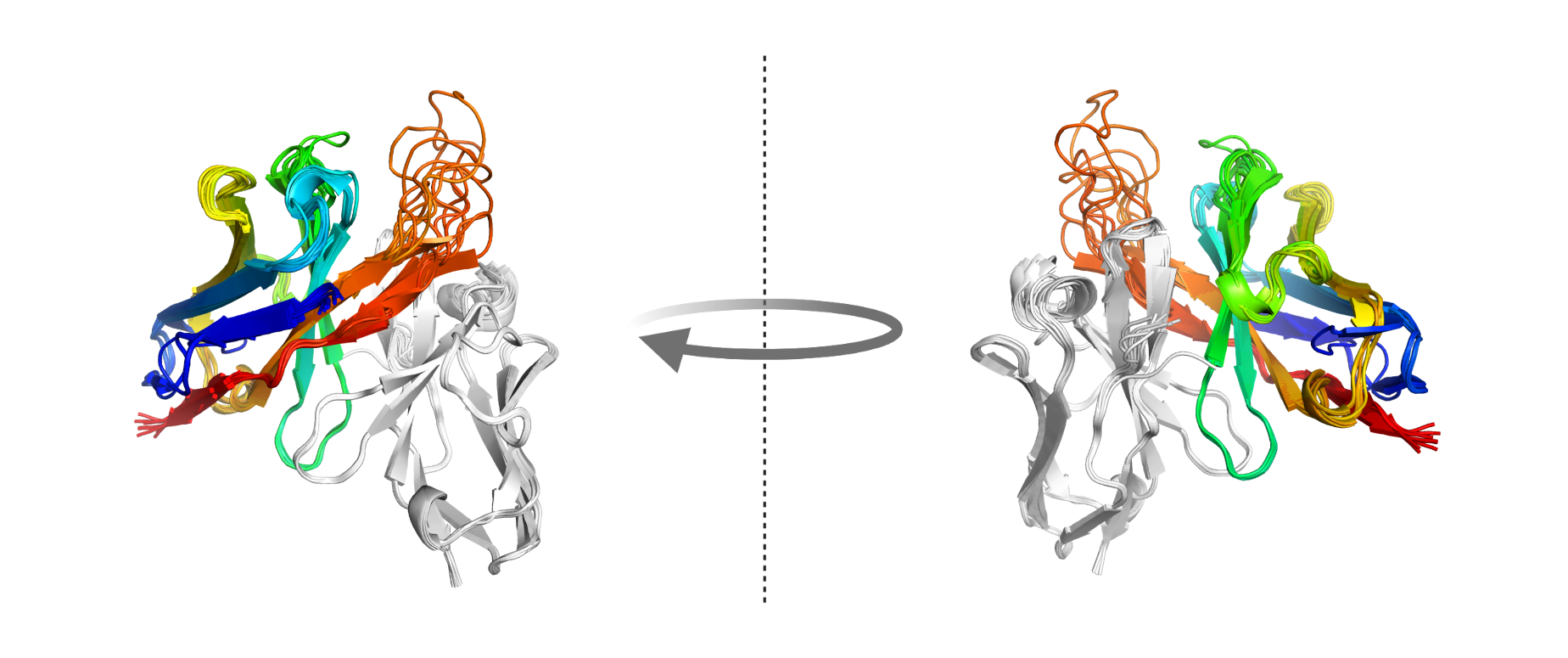}
  \caption{Structural models of example variable regions (Fv) with eight generated heavy chains given one input light chain. The generated heavy chains are colored in rainbow and the light chains are white.}
  \label{fig:generated_structures}
\end{figure}

We use clustalw \cite{Larkin2007Clustal2.0} in Biopython \cite{Cock2009Biopython:Bioinformatics} with default parameters to generate alignment profiles. Conservation analysis is generated by psiblast \cite{Altschup1990BasicTool} in Biopython onto UniRef90 database \cite{Suzek2007UniRef:Clusters}. To compare model confidence and sequence conservation, we apply softmax to PSSM and compare with the probability in next-word prediction. We use Logomaker \cite{Tareen2019Logomaker:Python} for visualization of sequence and alignment profiles. CDR and framework regions are defined in aho antibody renumbering scheme. CDRs of light chains are from residue ID 32 to 42, 57 to 76, and 109 to 138 for CDR L1, L2, and L3 respectively. CDRs of heavy chains are located from residue ID 24 to 42, 58 to 72, and 107 to 138.

\begin{figure}[H]
  \centering
  \includegraphics[width=0.6\linewidth]{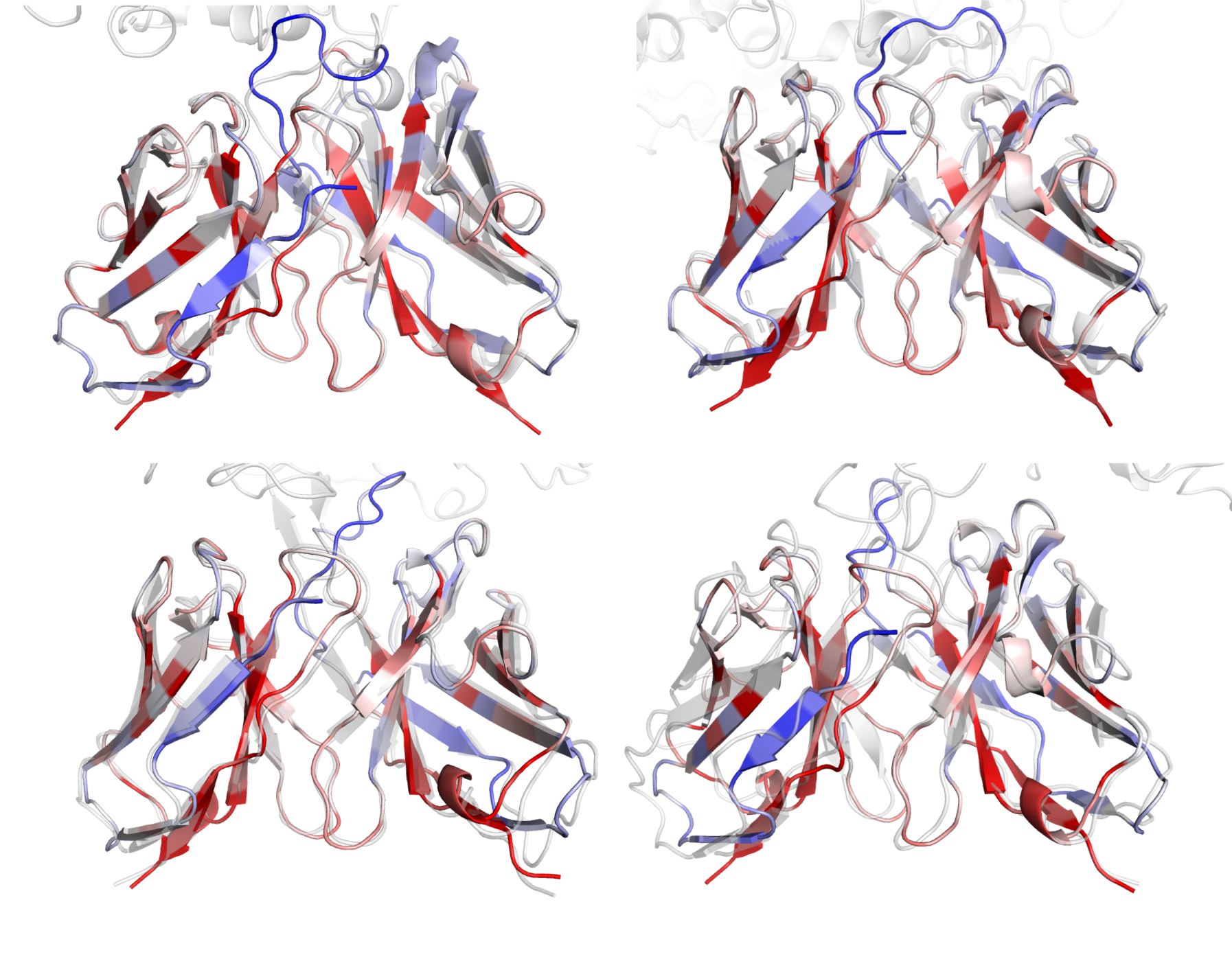}
  \caption{Structural overlay of capped average cross-attention from pairing partner onto each residue of SARS-Cov2-binding antibodies. Red color indicates regions highly attended while blue is weakly attended areas. (Upper right) PDB 6WPT. (Lower Left) PDB 7TB8 chain D and E. (Lower Right) PDB 7TB8 chain H and I. Consistently for all PDB structures, the CDR loops receive the least attention. This is consistent with the random nature of CDR loop sequences.}
  \label{fig:capped_cross_attentions}
\end{figure}

\begin{figure}[H]
  \centering
  \includegraphics[width=0.6\linewidth]{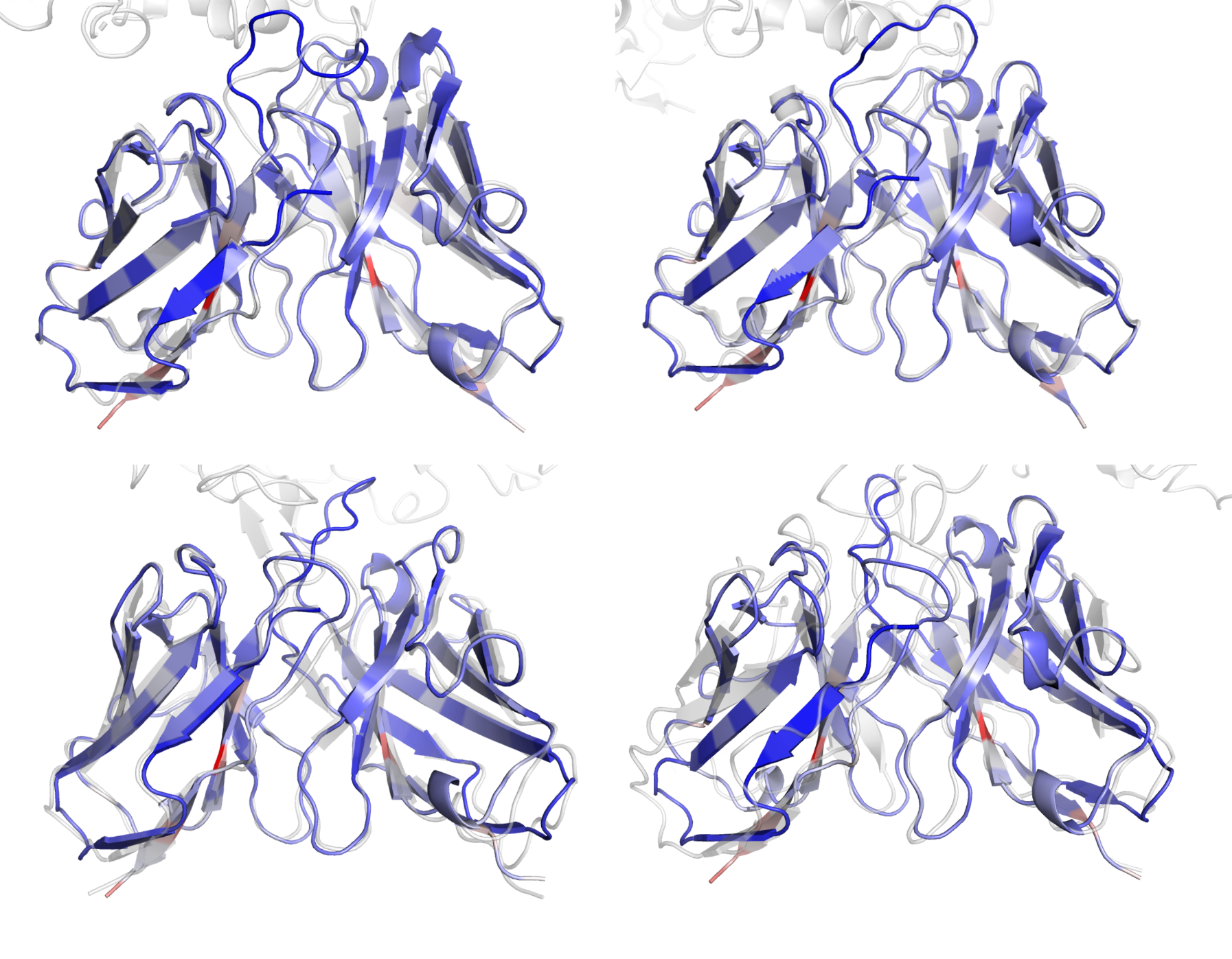}
  \caption{Structural overlay of uncapped average cross-attention from pairing partner onto each residue of SARS-Cov2-binding antibodies. (Upper right) PDB 6WPT. (Lower Left) PDB 7TB8 chain D and E. (Lower Right) PDB 7TB8 chain H and I.}
  \label{fig:uncapped_cross_attentions}
\end{figure}

\begin{figure}[H]
  \centering
  \includegraphics[width=0.6\linewidth]{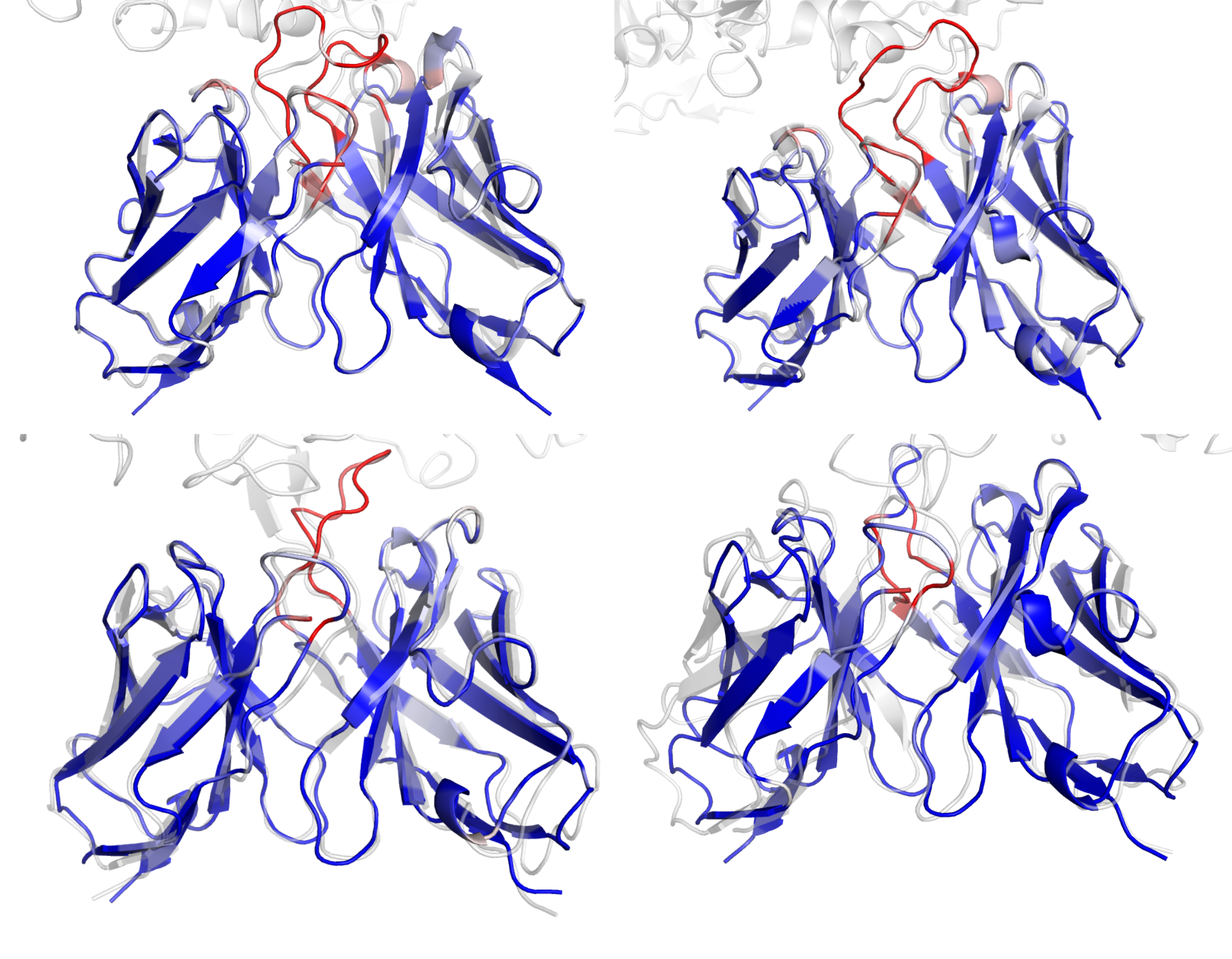}
  \caption{Structural overlay of capped next word prediction entropy of SARS-Cov2-binding antibodies. (Upper left) PDB 6WPS. (Upper right) PDB 6WPT. (Lower Left) PDB 7TB8 chain D and E. (Lower Right) PDB 7TB8 chain H and I.}
  \label{fig:capped_entropy}
\end{figure}

\begin{figure}[H]
  \centering
  \includegraphics[width=0.6\linewidth]{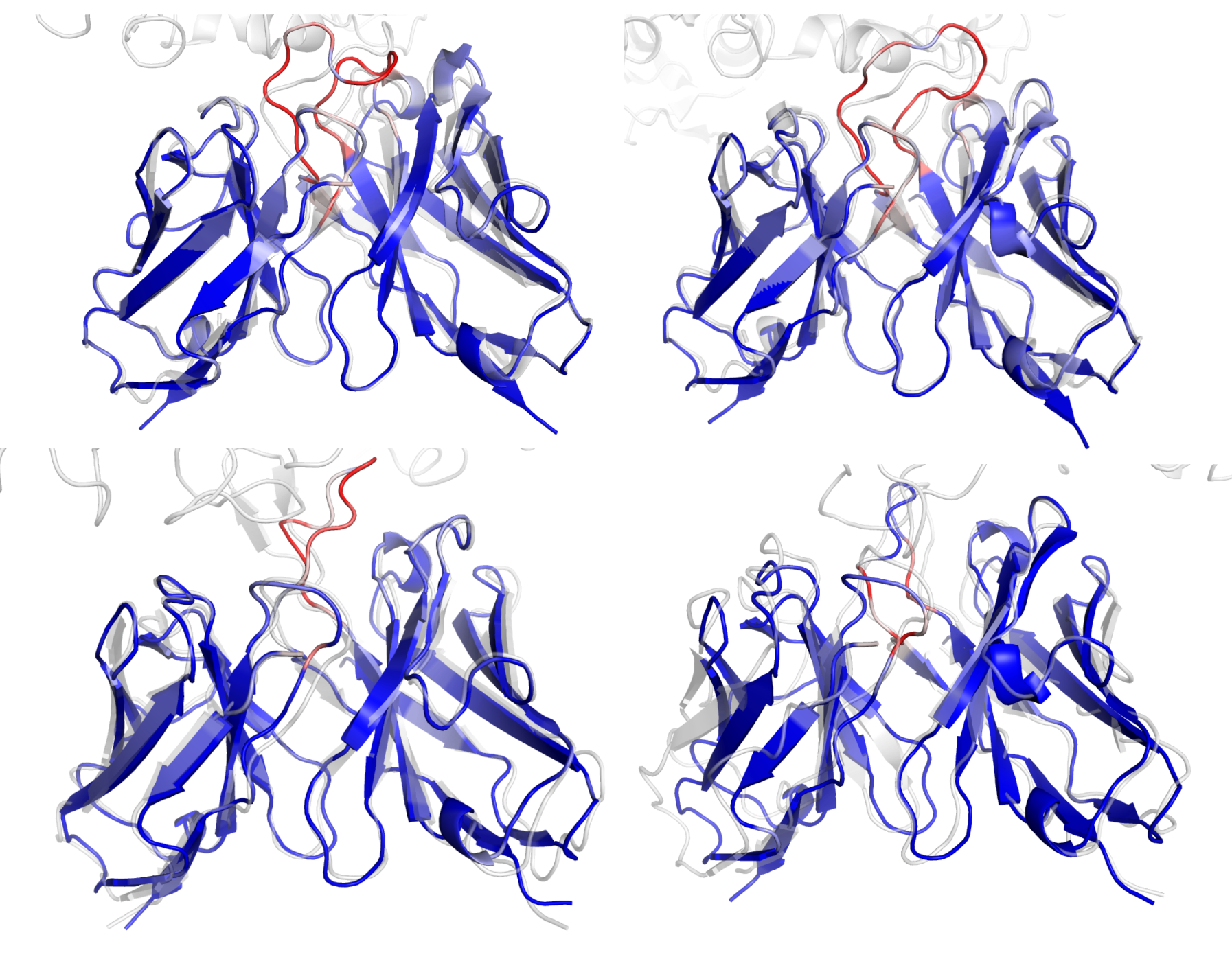}
  \caption{Structural overlay of uncapped next word prediction entropy of SARS-Cov2-binding antibodies. (Upper left) PDB 6WPS. (Upper right) PDB 6WPT. (Lower Left) PDB 7TB8 chain D and E. (Lower Right) PDB 7TB8 chain H and I.}
  \label{fig:uncapped_entropy}
\end{figure}

We overlay entropy and cross-attention per query residue onto antibody structures in PyMOL \cite{PyMOL}. Structural models are generated from DeepAb \cite{Ruffolo2022AntibodyLearning}, and in the case with available crystal structures, we align the models to the crystal chains to standardize numbering and fill in missing residues. We cap the values of average entropy and cross-attention per query residue in structural overlay and normalize heavy and light chains together for visualization purposes. Detailed visualization of capped and uncapped figures are also available (Figure \ref{fig:capped_cross_attentions}, \ref{fig:uncapped_cross_attentions}, \ref{fig:capped_entropy}, and \ref{fig:uncapped_entropy}).

\begin{table}[H]
  \centering
  \begin{tabular}{ccc}
    \toprule
    Region & Light & Heavy \\
    \midrule
    FR1 & 0.57$\pm$0.18 & 0.63$\pm$0.21 \\
    CDR1 & 0.36$\pm$0.26 & 0.41$\pm$0.22 \\
    FR2 & 0.77$\pm$0.13 & 0.76$\pm$0.14 \\
    CDR2 & 0.38$\pm$0.21 & 0.41$\pm$0.19 \\
    FR3 & 0.71$\pm$0.12 & 0.63$\pm$0.18 \\
    CDR3 & 0.31$\pm$0.20 & 0.22$\pm$0.14 \\
    FR4 & 0.76$\pm$0.18 & 0.90$\pm$0.09 \\
    whole sequence & 0.60$\pm$0.13 & 0.59$\pm$0.14 \\
    \bottomrule
  \end{tabular}
  \vspace{2mm}
  \caption{Sequence identities between generated and target sequences in test set by regions and target chain type.}
  \label{tab:generation_identity_regions}
\end{table}

\begin{table}[h]
\centering
{\tabulinesep=1.25mm
\begin{tabu}{c c c c c}
  \hline
  \multirow{2}{*}{Region} & \multicolumn{2}{c}{Light} & \multicolumn{2}{c}{Heavy} \\
  \cline{2-5} & Observed & Generated & Observed & Generated \\
  \hline
  FR1 & 22.75$\pm$0.43 & 22.74$\pm$0.44 & 28.91$\pm$1.45 & 28.99$\pm$0.06 \\
  FR2 & 15.00$\pm$0.00 & 15.00$\pm$0.00 & 14.00$\pm$0.00 & 14.00$\pm$0.00 \\
  FR3 & 32.02$\pm$0.20 & 32.00$\pm$0.04 & 32.00$\pm$0.00 & 32.00$\pm$0.00 \\
  FR4 & 9.97$\pm$0.22 & 10.00$\pm$0.03 & 11.00$\pm$0.00 & 10.96$\pm$0.33 \\
  CDR1 & 12.50$\pm$2.16 & 12.54$\pm$2.14 & 6.32$\pm$0.75 & 6.22$\pm$0.62 \\
  CDR2 & 7.03$\pm$0.38 & 7.02$\pm$0.25 & 16.80$\pm$0.77 & 16.82$\pm$0.66 \\
  CDR3 & 9.24$\pm$0.96 & 9.44$\pm$1.01 & 11.47$\pm$4.00 & 12.29$\pm$4.16 \\
  whole sequence & 108.51$\pm$2.38 & 108.74$\pm$2.29 & 120.45$\pm$4.57 & 121.28$\pm$4.18 \\
  \hline
\end{tabu}}
\vspace{2mm}
\caption{Sequence length of observed and generated sequences in test set by regions and chain type.}
\label{tab:generation_length_regions}
\end{table}

\begin{table}[H]
  \centering
  \begin{tabular}{ccc}
    \toprule
              & Heavy chain target & Light chain target \\
    \midrule
    Human     & 0.61$\pm$0.14 & 0.60$\pm$0.14 \\
    Mouse     & 0.56$\pm$0.10 & 0.62$\pm$0.10 \\
    \bottomrule
  \end{tabular}
  \vspace{2mm}
  \caption{Sequence identities between generated and target sequences in test set by species and target chain type}
  \label{tab:generation_identity}
\end{table}

\begin{sidewaysfigure}
  \centering
  \includegraphics[width=\textwidth]{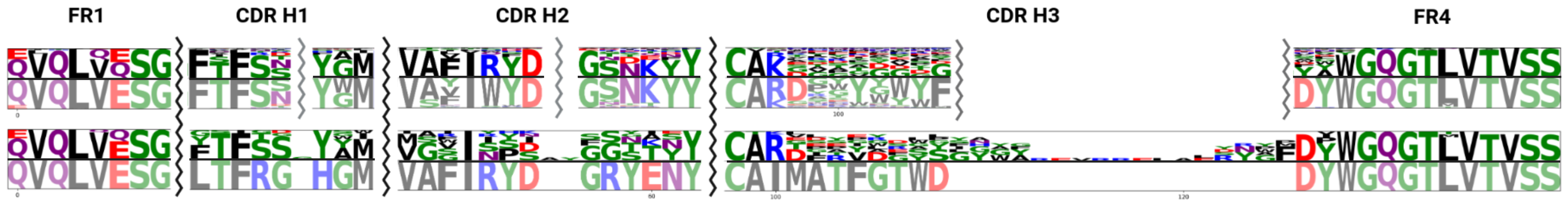}
  \caption{Comparison between observed and modeled alignment profiles on heavy chain in framework regions (FRs) CDR loops. (First row) Next-word probability in teacher-forcing. (Second row) Sequence conservation from position-specific scoring matrix. (Third row) Global alignment of generated sequences to (fourth row) the observed sequence. In general, generated sequences are more variable than next-word probability due to the cascade effect in iterative sampling, and might have different gene locus and/or families from the target sequence. The full-length alignment profiles of heavy and light chains together with four other output examples randomly from test set are available in Figure \ref{fig:alignment_profile_full}, \ref{fig:alignment_profile_full_light} and \ref{fig:alignment_profile_examples}.}
\end{sidewaysfigure}

\begin{sidewaysfigure}
  \centering
  \includegraphics[width=\textwidth]{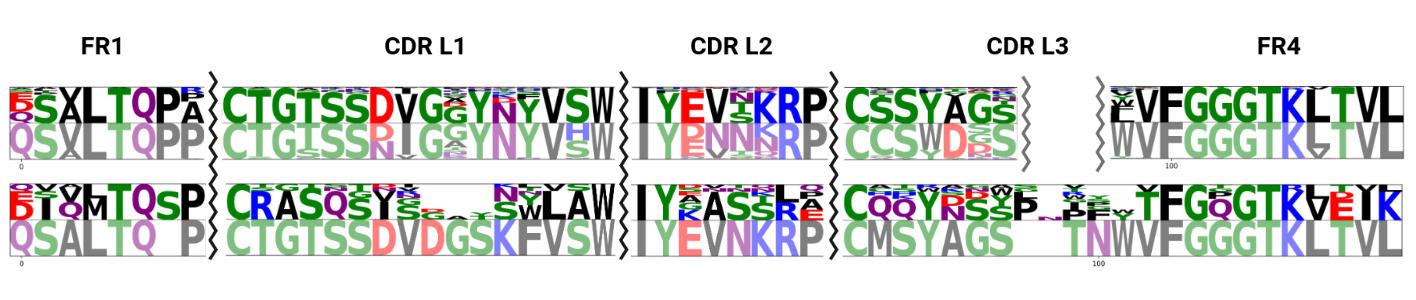}
  \caption{Comparison between observed and modeled alignment profiles on heavy chain in framework regions (FRs) CDR loops. (First row) Next word prediction probability. (Second row) Sequence conservation from position-specific scoring matrix. (Third row) Global alignment of generated sequences to (fourth row) the observed sequence. The heavy chain in Figure \ref{fig:alignment_profile} and the light chain here originate from the same observed antibody chain pair.}
  \label{fig:alignment_profile_light}
\end{sidewaysfigure}

\begin{sidewaysfigure}
  \centering
  \begin{subfigure}{\textwidth}
    \includegraphics[width=\textwidth]{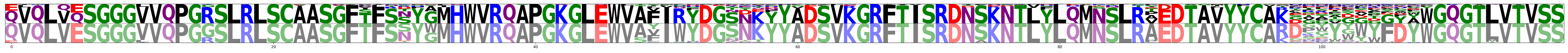}
    \caption{Next word prediction probability (top) versus position-specific scoring matrix (bottom)}
  \end{subfigure}
  \begin{subfigure}{\textwidth}
    \includegraphics[width=\textwidth]{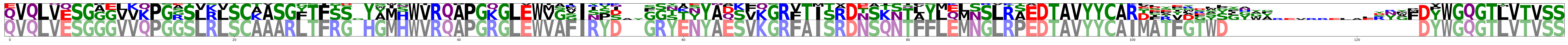}
    \caption{Generated sequences (top) versus observed sequence (bottom)}
  \end{subfigure}
  \caption{Full-length alignment profile of heavy chain between model predictions, conservation profile and observed sequence in Figure \ref{fig:alignment_profile}.}
  \label{fig:alignment_profile_full}
\end{sidewaysfigure}

\begin{sidewaysfigure}
  \centering
  \begin{subfigure}{\textwidth}
    \includegraphics[width=\textwidth]{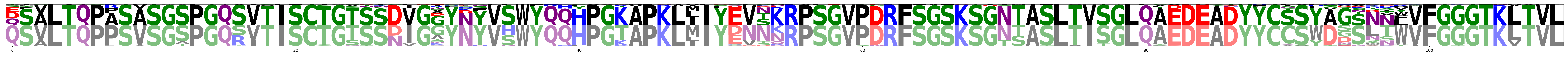}
    \caption{Next word prediction probability (top) versus position-specific scoring matrix (bottom)}
  \end{subfigure}
  \begin{subfigure}{\textwidth}
    \includegraphics[width=\textwidth]{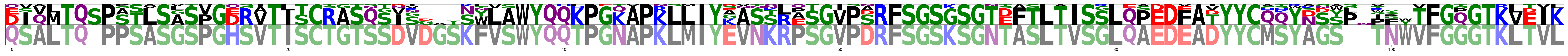}
    \caption{Generated sequences (top) versus observed sequence (bottom)}
  \end{subfigure}
  \caption{Full-length alignment profile of light chain between model predictions, conservation profile and observed sequence in Figure \ref{fig:alignment_profile_light}. }
  \label{fig:alignment_profile_full_light}
\end{sidewaysfigure}

\begin{sidewaysfigure}
  \centering
  \begin{subfigure}{\textwidth}
    \includegraphics[width=\textwidth]{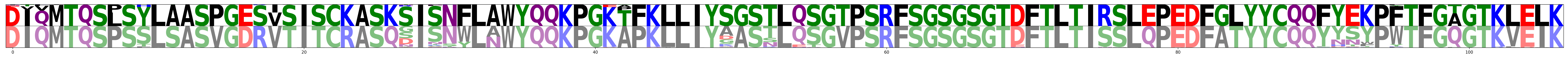}
    \caption{Next word prediction probability (top) versus position-specific scoring matrix (bottom)}
  \end{subfigure}
  \begin{subfigure}{\textwidth}
    \includegraphics[width=\textwidth]{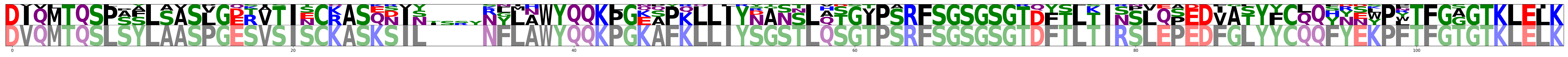}
    \caption{Generated sequences (top) versus observed sequence (bottom)}
  \end{subfigure}
  \begin{subfigure}{\textwidth}
    \includegraphics[width=\textwidth]{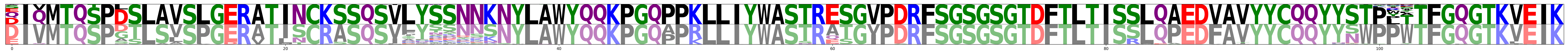}
    \caption{Next word prediction probability (top) versus position-specific scoring matrix (bottom)}
  \end{subfigure}
  \begin{subfigure}{\textwidth}
    \includegraphics[width=\textwidth]{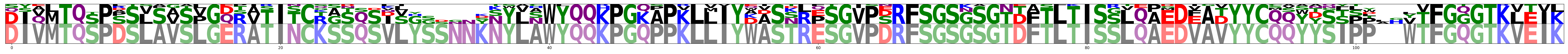}
    \caption{Generated sequences (top) versus observed sequence (bottom)}
  \end{subfigure}
    \begin{subfigure}{\textwidth}
    \includegraphics[width=\textwidth]{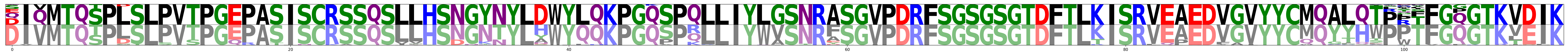}
    \caption{Next word prediction probability (top) versus position-specific scoring matrix (bottom)}
  \end{subfigure}
  \begin{subfigure}{\textwidth}
    \includegraphics[width=\textwidth]{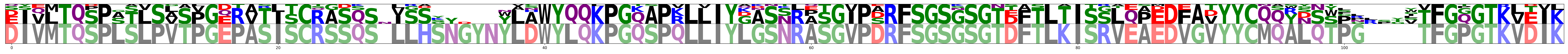}
    \caption{Generated sequences (top) versus observed sequence (bottom)}
  \end{subfigure}
    \begin{subfigure}{\textwidth}
    \includegraphics[width=\textwidth]{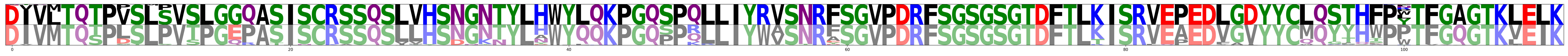}
    \caption{Next word prediction probability (top) versus position-specific scoring matrix (bottom)}
  \end{subfigure}
  \begin{subfigure}{\textwidth}
    \includegraphics[width=\textwidth]{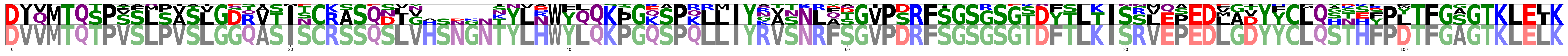}
    \caption{Generated sequences (top) versus observed sequence (bottom)}
  \end{subfigure}
  \caption{Four other examples of full-length alignment profile of light chain between model predictions, conservation profile and observed sequence. Examples are randomly drawn from all test set translations.}
  \label{fig:alignment_profile_examples}
\end{sidewaysfigure}

\begin{figure}[H]
  \centering
  \includegraphics[width=\linewidth]{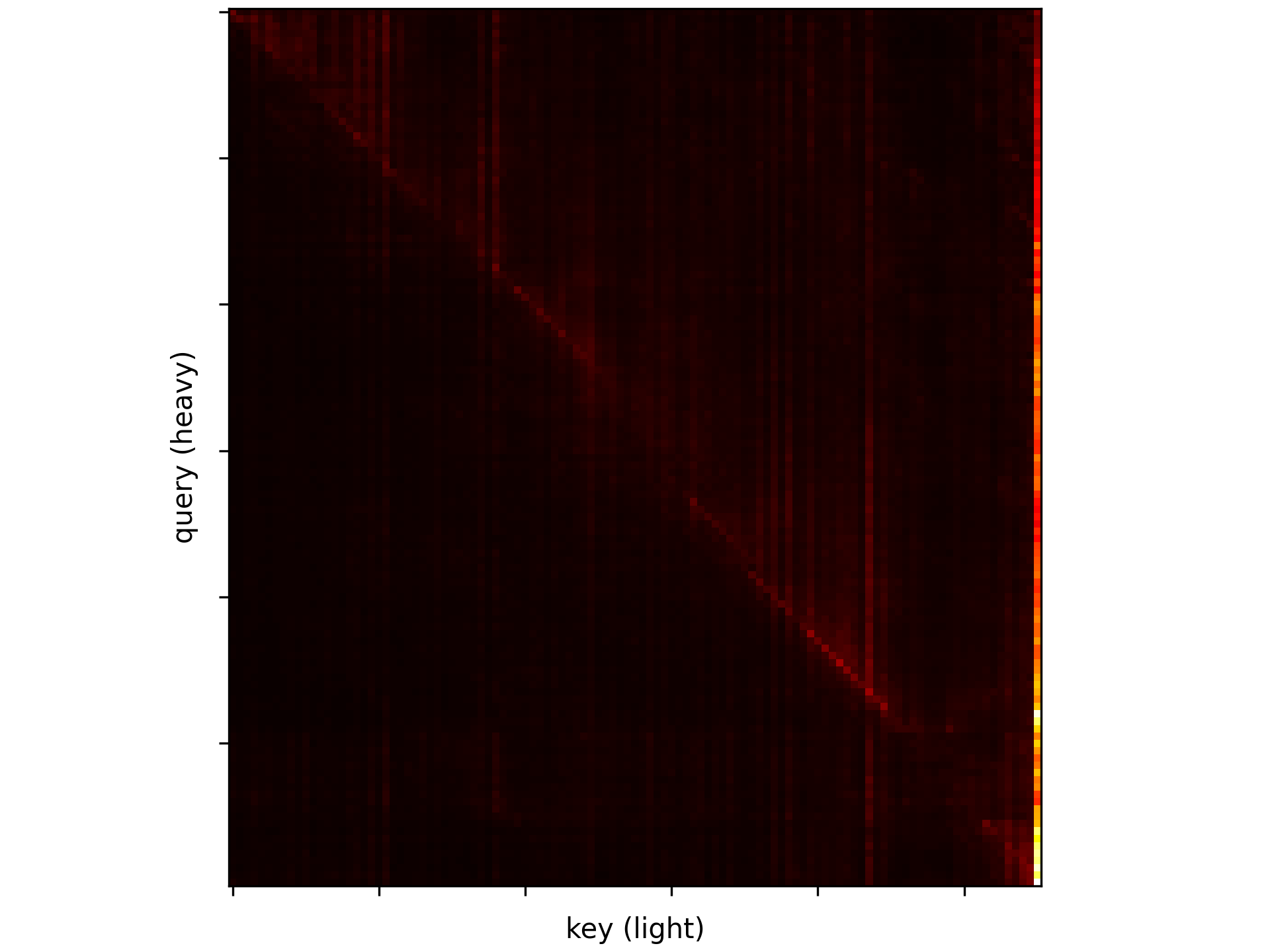}
  \caption{Cross-attention map between target heavy chain and input light chain in Figure \ref{fig:alignment_profile} averaged throughout heads and layers. Hypervariable regions generally receive less attention from queries consistently throughout all paired antibodies in the test set.}
\end{figure}

\subsection{Zero-shot Prediction from Paired Antibody Perplexity}
The emergence of protein function prediction from sequences alone can be traced back to conservation analysis. The idea is that residues detrimental to the function(s) of the protein should be conserved while other positions have more freedom to vary. Encoder-only protein LMs were shown to generalize Pott’s model \cite{Dauparas2019UnifiedSequences}, and outperform positional-specific scoring matrix (PSSM) with zero-shot prediction \cite{Meier2021LanguageFunction}. Similarly, the perplexity of decoder-only models is found to correlate with unseen experiment measurements \cite{Nijkamp2022ProGen2:Models}, while the same log-likelihood analysis can also be replicated on conditional sequence generation in inverse folding \cite{Hsu2022LearningStructures,Yang2022MaskedLearning}. Zero-shot and few-shot predictions from language pretraining are not unique to protein LMs but arise generally from large-scale language modeling.

\begin{figure}[h]
  \centering
  \includegraphics[width=\linewidth]{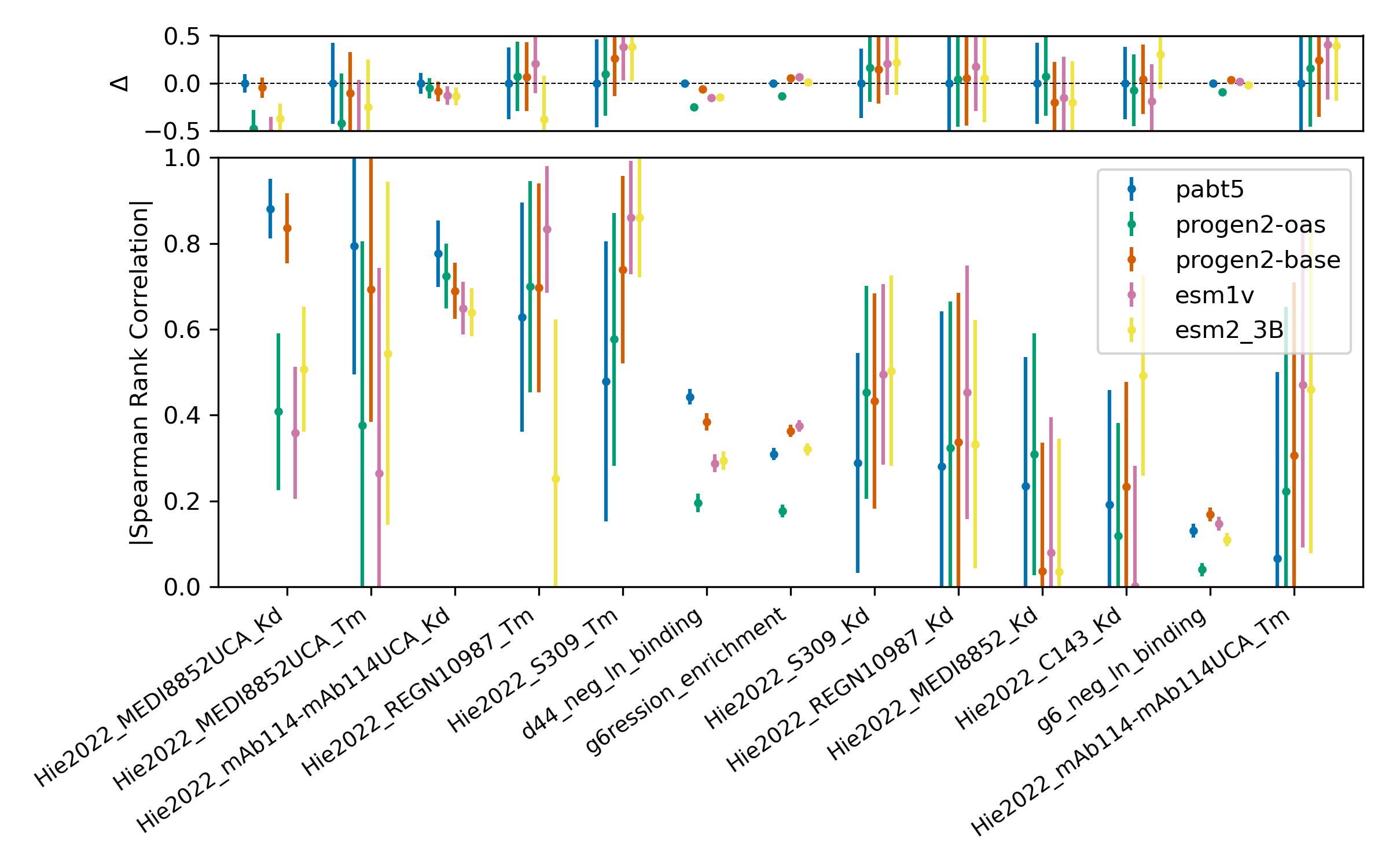}
  \caption{Zero-shot prediction performance on antibody measurements of our model and state-of-the-art. x-axis represents the antibody functional datasets. (Top) The difference in absolute spearman rank correlation (SRC) between our model and state-of-the-art. (Bottom) Absolute SRC between model (pseudo-)perplexity and measurements. Error bars are estimated in standard deviation with 1000 bootstrap samples.}
  \label{fig:sota_zeroshot}
\end{figure}

Benchmarked on antibody functional datasets, we show that our model has competitive results with the current state-of-the-art protein LMs. We benchmark our model on 13 antibody functional datasets on either stability, binding affinity or expression measurements \cite{Koenig2017MutationalBinding,Warszawski2019OptimizingInterfaces,Hie2022EfficientAlone} in Figure \ref{fig:sota_zeroshot}. Our encoder-decoder model achieves a similar performance as ProGen2 and is better than ProGen2-OAS, which is finetuned on the unpaired OAS dataset. The major architectural difference is that ProGen2 is a decoder-only model which requires joining heavy and light chain sequences with a GS linker, whereas our encoder-decoder model computes the average perplexity of forward- and back-translations. Nonetheless, ProGen2 and ProGen2-OAS have fewer parameters than our model, making model comparison difficult. In addition, we have also included pseudo-perplexity from encoder-only models (ESM) \cite{Meier2021LanguageFunction,Lin2022LanguagePrediction} to highlight the difference in architecture.

\begin{figure}[H]
  \centering
  \includegraphics[width=0.75\linewidth]{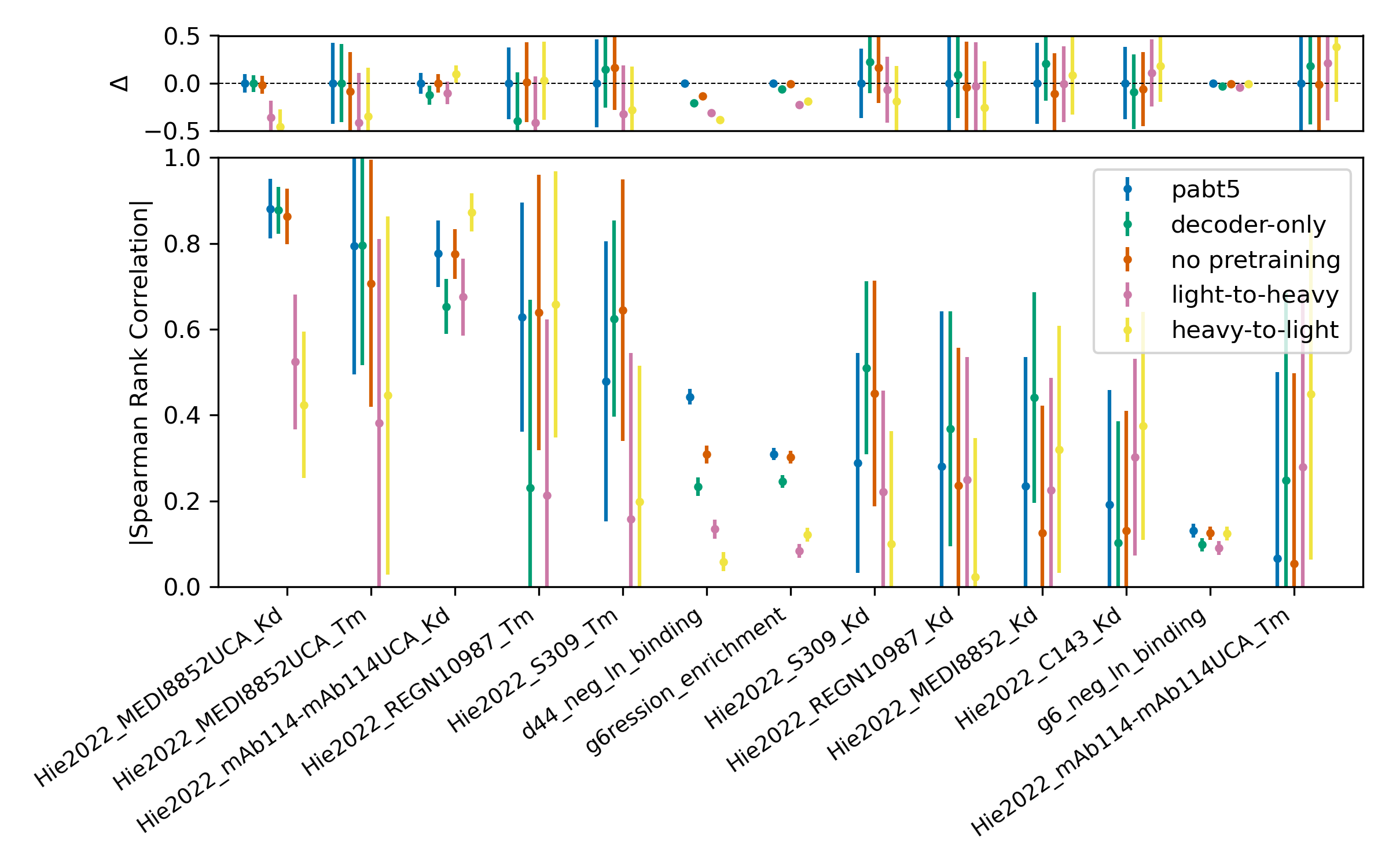}
  \caption{Ablation study on zero-shot prediction on all datasets. x-axis represents datasets. (Top) The difference in absolute spearman rank correlation (SRC) between our model and ablation. (Bottom) Absolute SRC between model (pseudo-)perplexity and measurements. Error bars are estimated in standard deviation with 1000 bootstrap samples.}
  \label{fig:zeroshot_ablation_all}
\end{figure}

To further investigate the impact of each component in our model, we perform an ablation study on the need for an encoder-decoder architecture, bidirectional translations in evaluation, and pretraining. For any comparison with statistical significance (p-value $<$ 0.05), our encoder-decoder model always outperforms ablations (Figure \ref{fig:zeroshot_ablation_all}).

We evaluate the perplexity from the benchmarked models and calculate the absolute value of spearman rank correlation (SRC) with the experimental measurements. By default, we define a symmetric paired perplexity by taking the average of that in forward- and back-translations for zero-shot prediction. Since ProGen2 is a decoder-only model, we join the heavy and light chains by a GS linker of \textit{GGGGSGGGGSGGGGS} and parse the paired antibody as a single sequence. In the case of our decoder-only ablation, we train the model without an encoder but take the average of heavy and light chain perplexities. Our ablation on pretraining from ProtT5 shares the same hyperparameters in Section \ref{subsubsection:optimization}. The mean and standard deviation of SRC are estimated by bootstrapping 1000 samples.

\section{Sequence Clustering}
\label{appendix:cluster_split}
\setcounter{figure}{0}
\setcounter{table}{0}

Contrary to using all non-redundant sequences in the dataset, one can cluster these sequences by an identity cutoff and include only the representative sequences of each cluster. This provides a few advantages. First, it reduces the dataset size and increases sparsity for efficient training. Second, it de-biases the database from heavily studied families. Third, it provides a better assessment of model generalizability by limiting the information shared between train and test sets. 

This section investigates the impact of sequence clustering on paired OAS dataset and our model performance. We argue that for our specific case, including all non-redundant sequences helps the model in three ways. While sequence clustering affects the performance evaluation, the impact is minor and does not affect conclusions.
\begin{itemize}
\item Sequence clustering reduces the size of paired OAS dataset by at least 50\%.
\item Fine-grained resolution in a subspace of protein universe helps resolve all antibodies and their pairings, in particular for learning gene families.
\item De-biasing might fail to reflect the preference(s) of antibody pairing.
\end{itemize}

\subsection{Impact on Dataset Size}
We use linclust from mmseqs2 to cluster representative sequences with --min-seq-id to specify identity cutoff, and -c 0.8 and --cov-mode 1, and otherwise the default parameters. We do not observe any signs of truncation at the N- and C-termini on paired OAS dataset.

As reported in Table \ref{tab:dataset_size_id}, the dataset reduces in size exponentially with the identity threshold in clustering. For each increment of 5\%, the number of translations after clustering falls by about half. This impacts not only the training but also the statistical power of evaluation(s) given the size of the diminished test set.

From here, we denote exclusive node split in Section \ref{subsec:dataset_split} on clustered sequences as cluster split. We decide to repeat the analyses on cluster split with an identity cutoff of 95\% and compare with that from training on non-redundant sequences.

\begin{table}[H]
  \centering
  \begin{tabular}{ccccc}
    \toprule
    & non-redundant & 95\% & 90\% & 85\% \\
    \midrule
    Training set & 260062 & 127904 & 53814 & 22266 \\
    Validation set & 846 & 356 & 188 & 74 \\
    Test set & 802 & 346 & 178 & 78 \\
    \bottomrule
  \end{tabular}
  \vspace{2mm}
  \caption{Impact of identity threshold on dataset size in terms of number of translations}
  \label{tab:dataset_size_id}
\end{table}

\subsection{Impact on Results}
\subsubsection{Pairing Perplexity Reflects Preferences in Chain Pairing}

In double-random scheme, training and evaluation on clustered sequences result in higher accuracy in the first classification task but weaker in the second classification task. In both tasks, mispairing identification informed by model perplexity alone still outperforms the baseline. Similar observation holds also in single-random scheme \ref{tab:single-generation1_linclust} and \ref{tab:single-generation2_linclust}. Overall, the results are unaffected by sequence clustering.

\begin{table}[H]
\centering
\begin{minipage}{0.48\linewidth}
\centering
{\tabulinesep=1.1mm
\begin{tabu}{c c c}
  \multicolumn{3}{c}{\textbf{First Classification Task}} \\
  \hline
  Mispairing type & Target chain & Accuracy \\
  \hline
  \multirow{2}{*}{Chain type} & Light & 0.98 \\
  \cline{2-3} & Heavy & 0.98 \\
  \hline
  \multirow{2}*{Species} & Light & 0.85 \\
  \cline{2-3} & Heavy & 0.88 \\
  \hline
\end{tabu}}
\vspace{2mm}
\end{minipage}
\begin{minipage}{0.48\linewidth}
\centering
{\tabulinesep=1.1mm
\begin{tabu}{c c c}
  \multicolumn{3}{c}{\textbf{Second Classification Task}} \\
  \hline
  Mispairing type & Accuracy & AUROC \\
  \hline
  Chain type & 0.55 &  0.65\\
  Species & 0.55 &  0.57\\
  \hline
\end{tabu}}
\vspace{2mm}
\end{minipage}
\caption{Performance on first and second classification task on model perplexity alone. (Left) In the first classification task, mispairing assignment is based on the rank of perplexity without any parameterizable model. (Right) In the second classification task, instead of unidirectional translation, logistic regression is trained on the bidirectional average of translation perplexity in validation set, and evaluated on test set. Random assignment results in an accuracy of 0.5 in the first class, and an additional AUROC of 0.5 in the second task.}
\end{table}

\begin{table}[H]
\centering
{\tabulinesep=1.1mm
\begin{tabu}{c c c}
  \hline
  Mispairing type & Target chain & Accuracy \\
  \hline
  \multirow{2}{*}{Chain type} & Light &  0.92\\
  \cline{2-3} & Heavy & 1\\
  \hline
  \multirow{2}*{Species} & Light & 0.99\\
  \cline{2-3} & Heavy & 0.98\\
  \hline
\end{tabu}}
\vspace{2mm}
\caption{First classification task assignment accuracy by the perplexity rank between correct and \textit{mispaired} antibody sequences in single-generation scheme.}
\label{tab:single-generation1_linclust}
\end{table}

\begin{table}[H]
\centering
{\tabulinesep=1.1mm
\begin{tabu}{c c c}
  \hline
  Mispairing type & Accuracy & AUROC \\
  \hline
  Chain-type & 0.55 & 0.62\\
  Species & 0.56 & 0.62\\
  \hline
\end{tabu}}
\vspace{2mm}
\caption{Second classification task performance in single-generation scheme}
\label{tab:single-generation2_linclust}
\end{table}

\subsubsection{Sequence Generation Aligns with Conserved and Variable Domains in Antibodies}
Our model from cluster split still has high entropy and generates variable-length sequences at hypervariable domains. Results are largely unaffected by cluster split.

\begin{figure}[H]
  \centering
  \includegraphics[width=\textwidth]{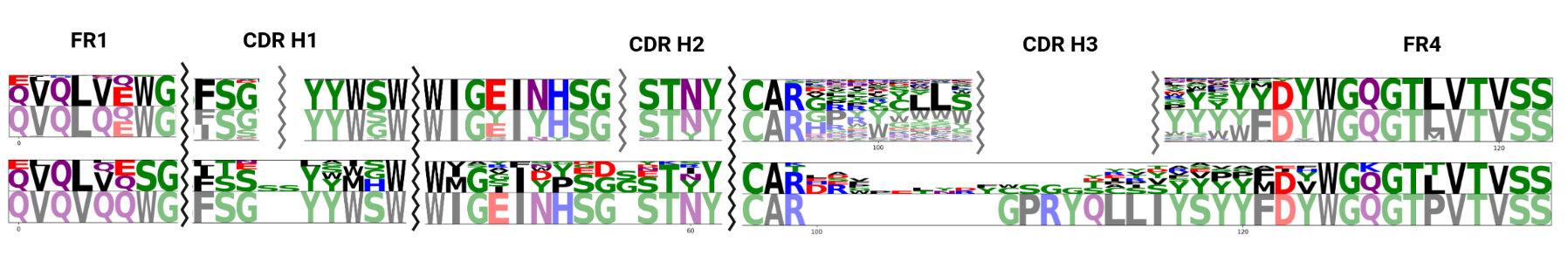}
  \caption{Comparison between observed and modeled alignment profiles on heavy chain in framework regions (FRs) CDR loops. (First row) Next word prediction probability. (Second row) Sequence conservation from position-specific scoring matrix. (Third row) Global alignment of generated sequences to (fourth row) the observed sequence.}
  \label{fig:alignment_profile_linclust}
\end{figure}

\begin{figure}[H]
  \centering
  \includegraphics[width=\linewidth]{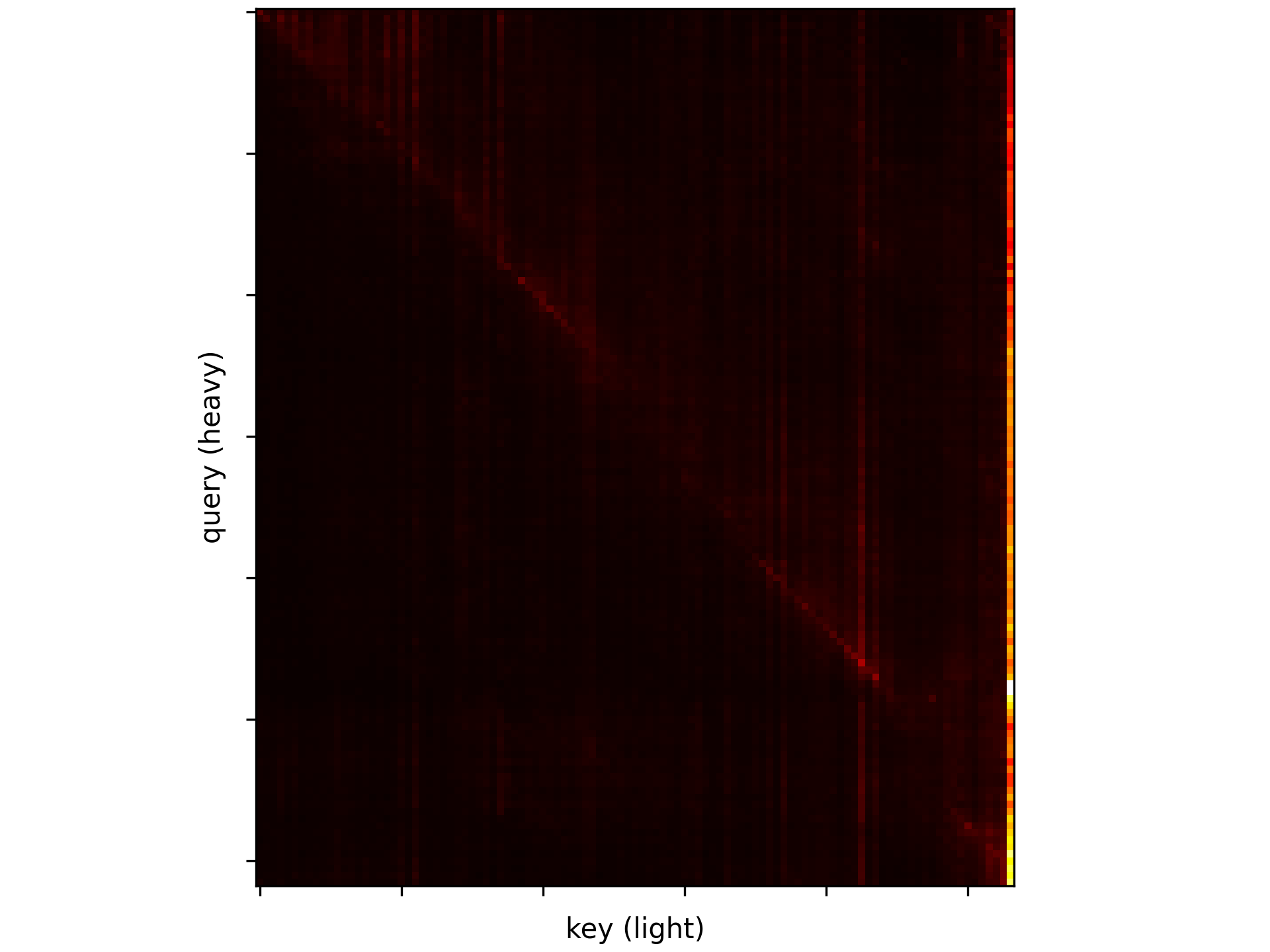}
  \caption{Cross-attention map between target heavy chain and input light chain in Figure \ref{fig:alignment_profile_linclust} averaged throughout heads and layers. Hypervariable regions generally receive less attention from queries.}
\end{figure}

\begin{table}[H]
  \centering
  \begin{tabular}{ccc}
    \toprule
    Region & Light & Heavy \\
    \midrule
    FR1 & 0.59$\pm$0.18 & 0.60$\pm$0.21 \\
    CDR1 & 0.35$\pm$0.25 & 0.38$\pm$0.20 \\
    FR2 & 0.78$\pm$0.13 & 0.76$\pm$0.13 \\
    CDR2 & 0.39$\pm$0.22 & 0.37$\pm$0.15 \\
    FR3 & 0.69$\pm$0.12 & 0.58$\pm$0.15 \\
    CDR3 & 0.33$\pm$0.20 & 0.24$\pm$0.15 \\
    FR4 & 0.79$\pm$0.19 & 0.90$\pm$0.08 \\
    whole sequence & 0.60$\pm$0.13 & 0.56$\pm$0.12 \\
    \bottomrule
  \end{tabular}
  \vspace{2mm}
  \caption{Sequence identities between generated and target sequences in test set by regions and target chain type.}
\end{table}

\begin{table}[H]
\centering
{\tabulinesep=1.25mm
\begin{tabu}{c c c c c}
  \hline
  \multirow{2}{*}{Region} & \multicolumn{2}{c}{Light} & \multicolumn{2}{c}{Heavy} \\
  \cline{2-5} & Observed & Generated & Observed & Generated \\
  \hline
  FR1 & 22.59$\pm$0.50 & 22.37$\pm$0.48 & 28.86$\pm$1.83 & 29.00$\pm$0.00 \\
  CDR1 & 12.74$\pm$2.18 & 12.87$\pm$1.41 & 6.26$\pm$0.67 & 6.03$\pm$0.25 \\
  FR2 & 15.00$\pm$0.00 & 15.00$\pm$0.00 & 14.00$\pm$0.00 & 14.00$\pm$0.00 \\
  CDR2 & 7.05$\pm$0.43 & 7.00$\pm$0.00 & 16.74$\pm$0.63 & 16.89$\pm$0.32 \\
  FR3 & 32.00$\pm$0.00 & 32.00$\pm$0.00 & 32.00$\pm$0.15 & 32.00$\pm$0.00 \\
  CDR3 & 9.63$\pm$1.06 & 9.95$\pm$0.81 & 12.32$\pm$3.93 & 16.51$\pm$3.97 \\
  FR4 & 9.94$\pm$0.36 & 10.00$\pm$0.00 & 11.00$\pm$0.00 & 11.00$\pm$0.00 \\
  whole sequence & 108.88$\pm$2.54 & 109.19$\pm$1.58 & 121.10$\pm$4.63 & 125.43$\pm$4.10 \\
  \hline
\end{tabu}}
\vspace{2mm}
\caption{Sequence length of observed and generated sequences in test set by regions and chain type.}
\end{table}

\subsubsection{Conditional Generation Recovers Pairing Sequences}
\begin{figure}[h]
  \centering
  \begin{subfigure}{0.3\textwidth}
    \includegraphics[width=\textwidth]{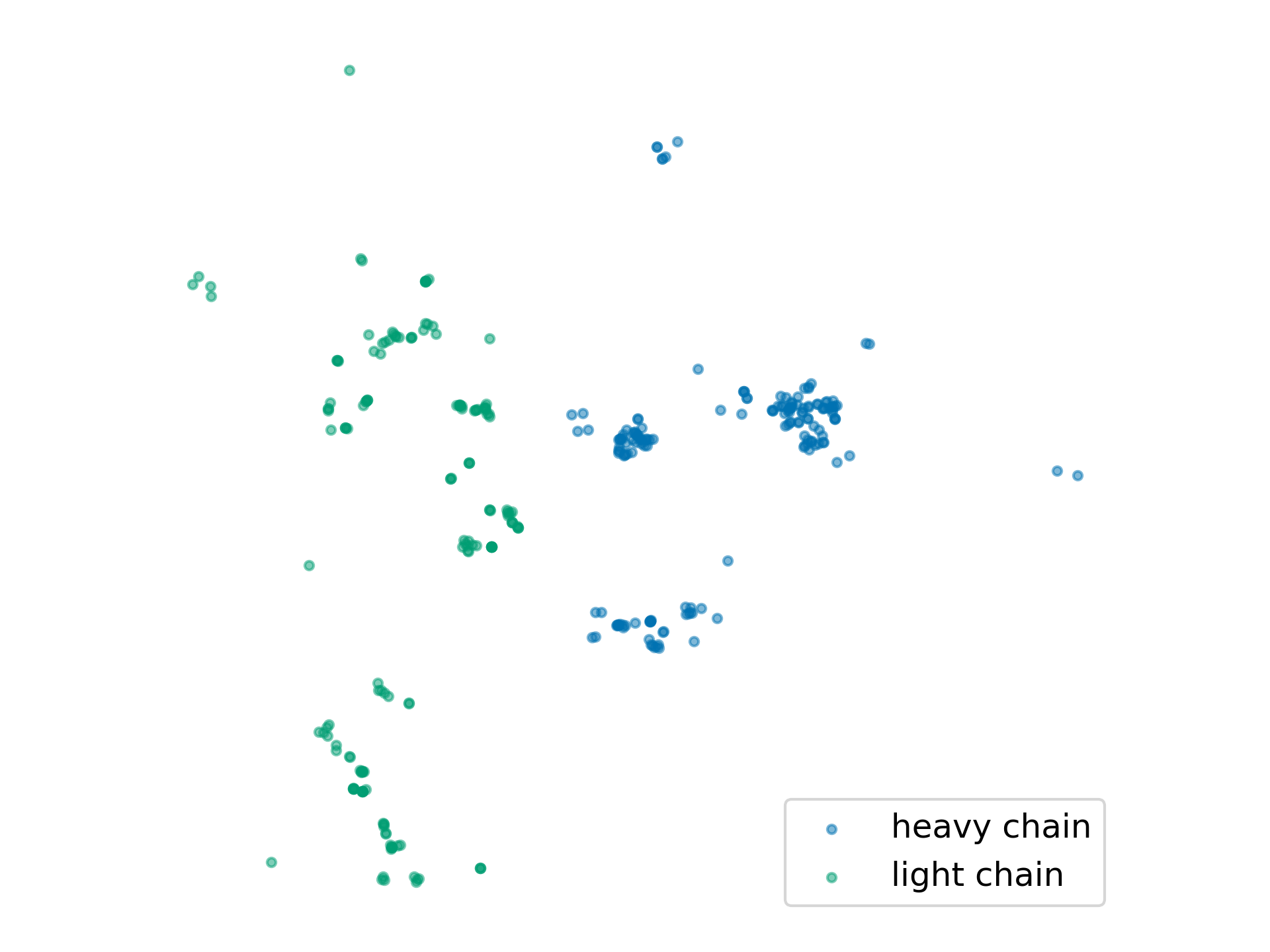}
    \caption{Heavy and light chains}
    \label{fig:tsne-heavy-light_linclust}
  \end{subfigure}
  \begin{subfigure}{0.3\textwidth}
    \includegraphics[width=\textwidth]{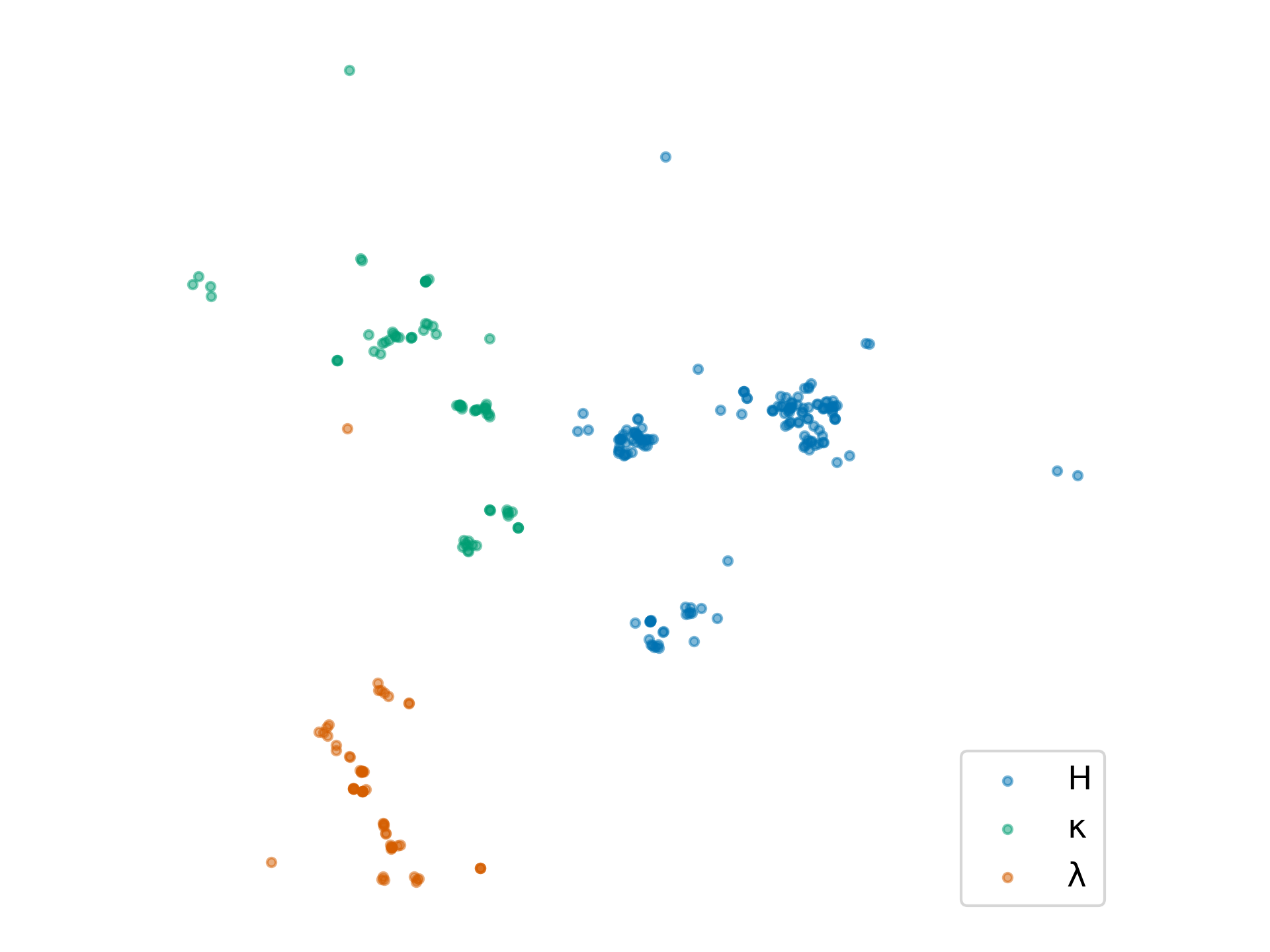}
    \caption{Gene loci in human}
    \label{fig:tsne-gene_loci_linclust}
  \end{subfigure}
  \begin{subfigure}{0.3\textwidth}
    \includegraphics[width=\textwidth]{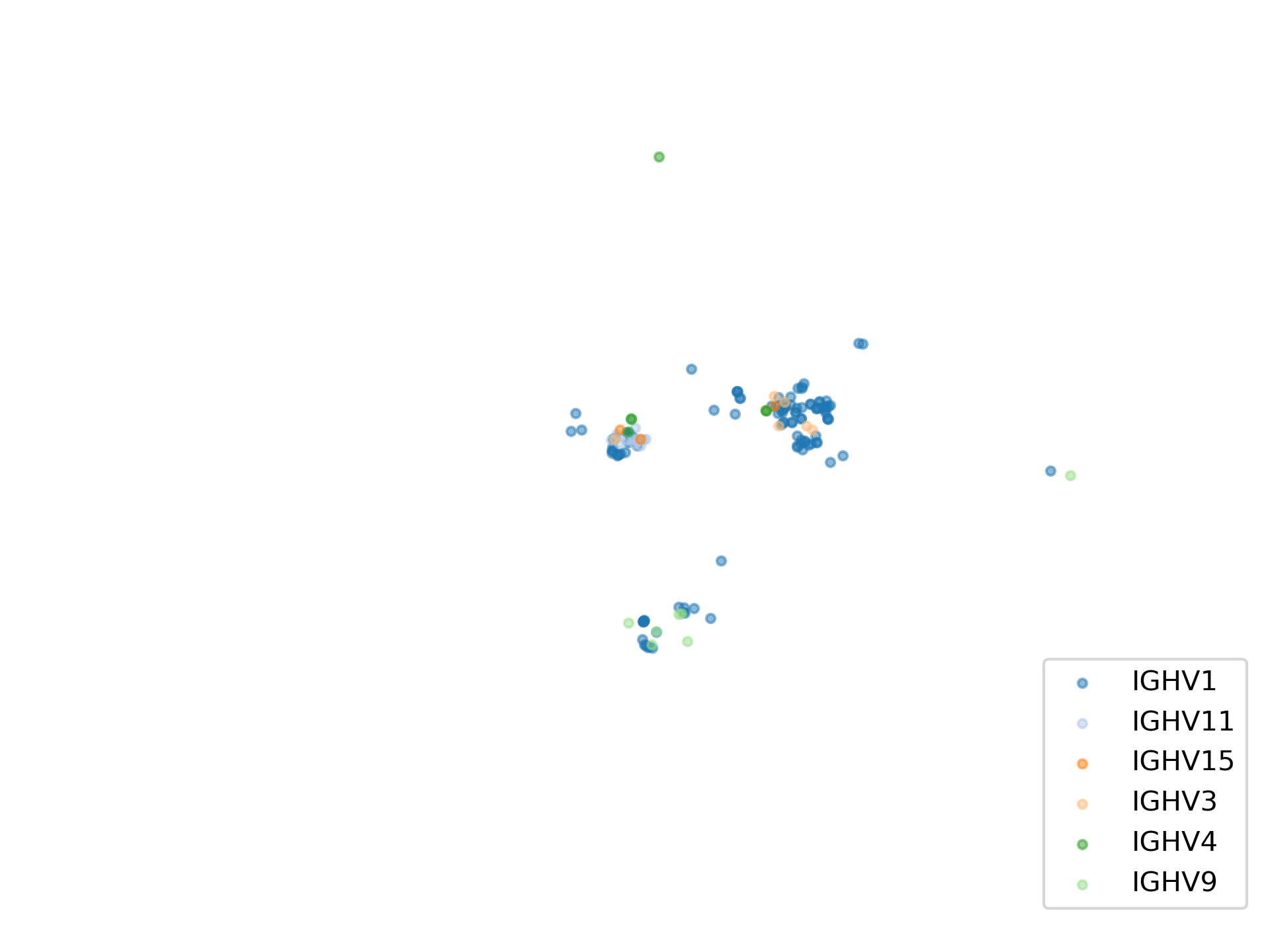}
    \caption{IGHV gene families in human}
    \label{fig:tsne-ighv_linclust}
  \end{subfigure}
  \caption{t-SNE plot of encoder hidden states of test set sequences in progressively fine categories (chain types, gene loci, and gene families).}
\end{figure}

\begin{figure}[H]
  \centering
  \includegraphics[width=0.7\linewidth]{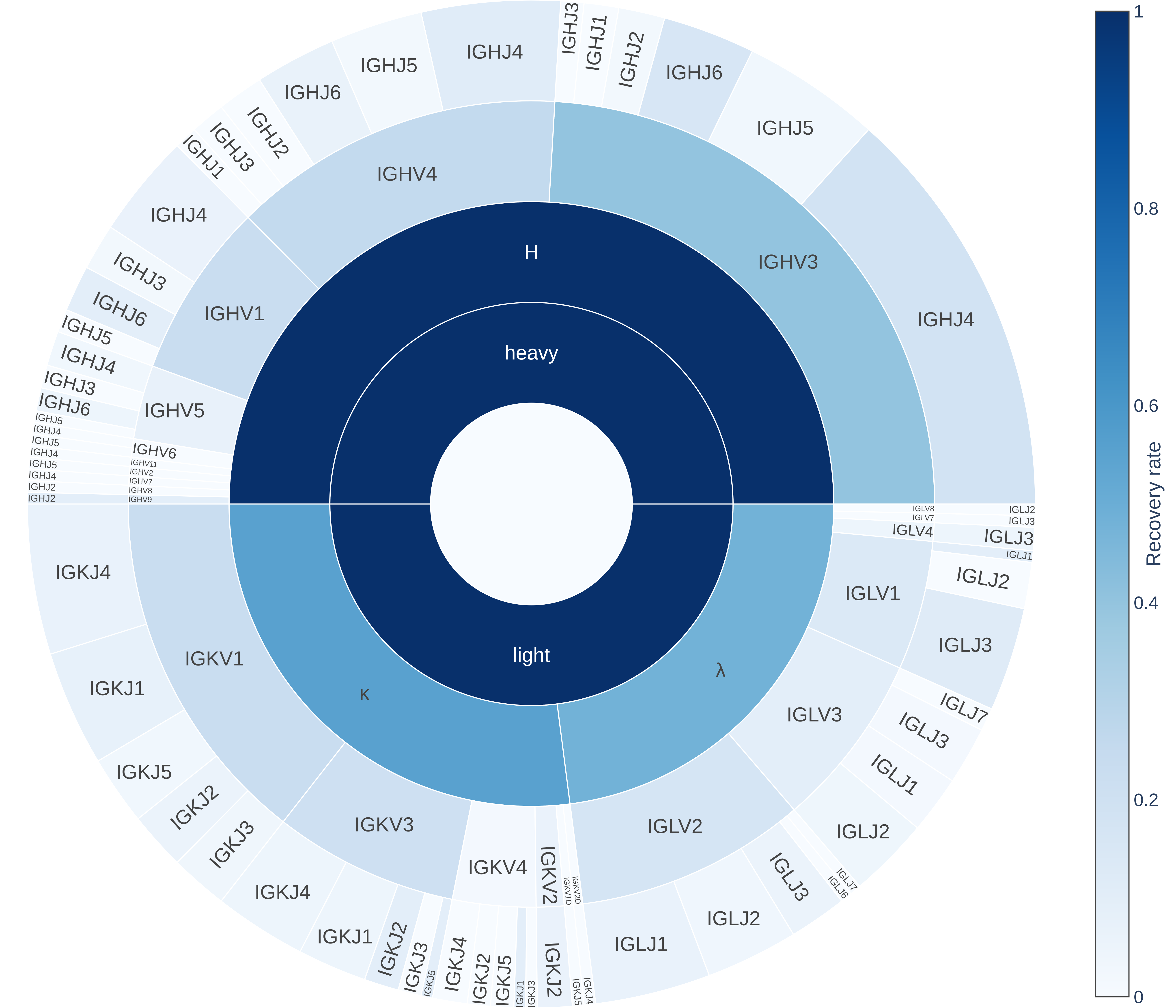}
  \caption{Recovery rate of target chain type, gene loci, and gene families in sequence generation. Performance is represented in a hierarchical order, where parent classes are centered while children categories are on the periphery. On each rim, the arc lengths of categories are proportional to their populations in test set. Dark blue represents perfect recovery whereas white color implies low recovery rate.}
  \label{fig:generation_sunbursts_linclust}
\end{figure}

\begin{figure}[H]
  \centering
  \includegraphics[width=0.75\linewidth]{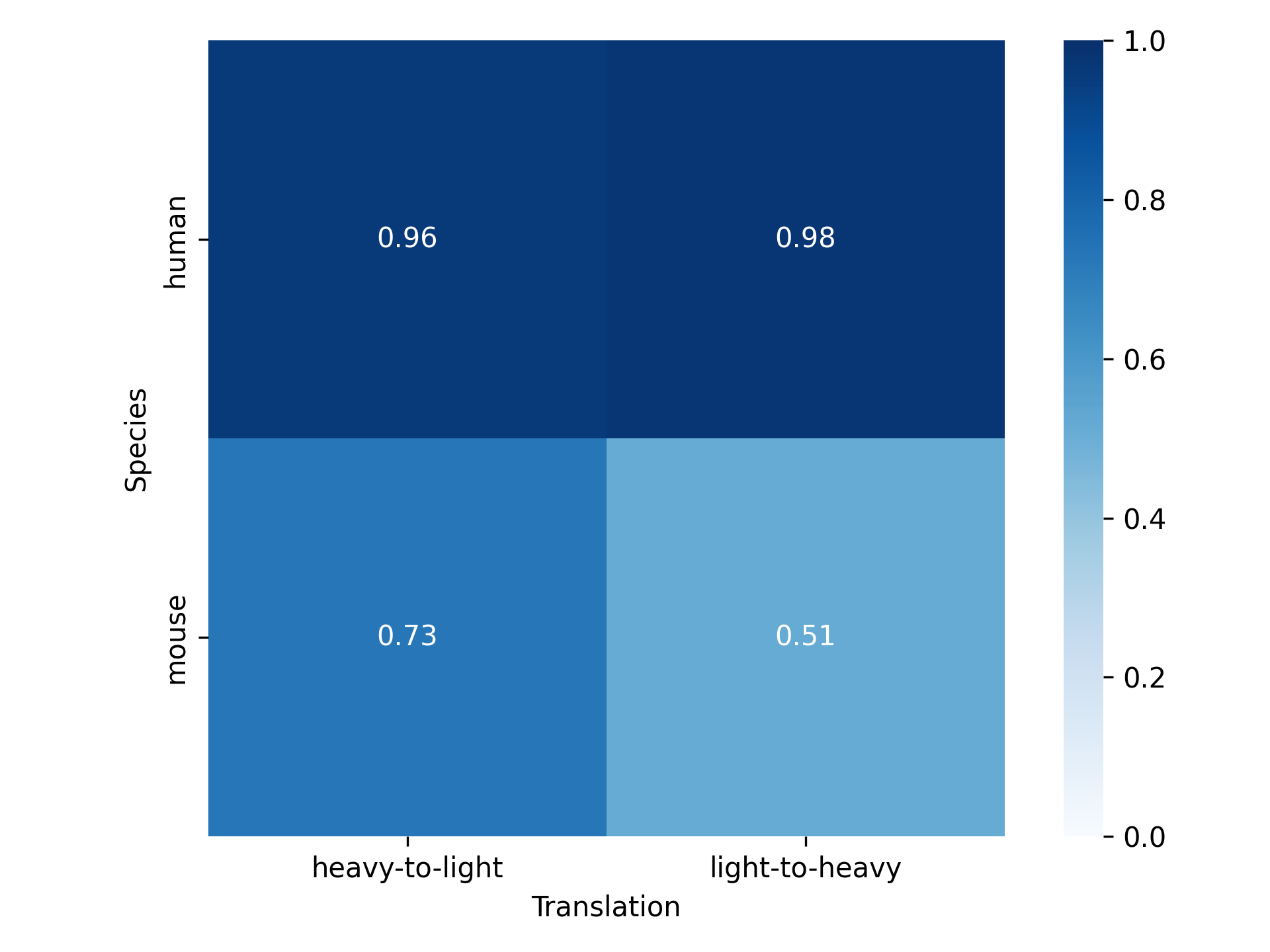}
  \caption{Recovery rate on species by original species and translation direction.}
  \label{fig:recovery_species_linclust}
\end{figure}

t-SNE plots on sequence representation are similar to those without sequence clustering (Figure \ref{fig:tsne-heavy-light_linclust}, \ref{fig:tsne-gene_loci_linclust} and \ref{fig:tsne-ighv_linclust}). When comparing on recovery rate of target sequences, we found that cluster split leads to slightly stronger bias towards specific families (Figure \ref{fig:generation_sunbursts_linclust}). Sequence recovery is similar to that without sequence clustering (Figure \ref{fig:recovery_species} and \ref{fig:recovery_species_linclust}).

\begin{figure}[H]
  \centering
  \includegraphics[width=0.5\linewidth]{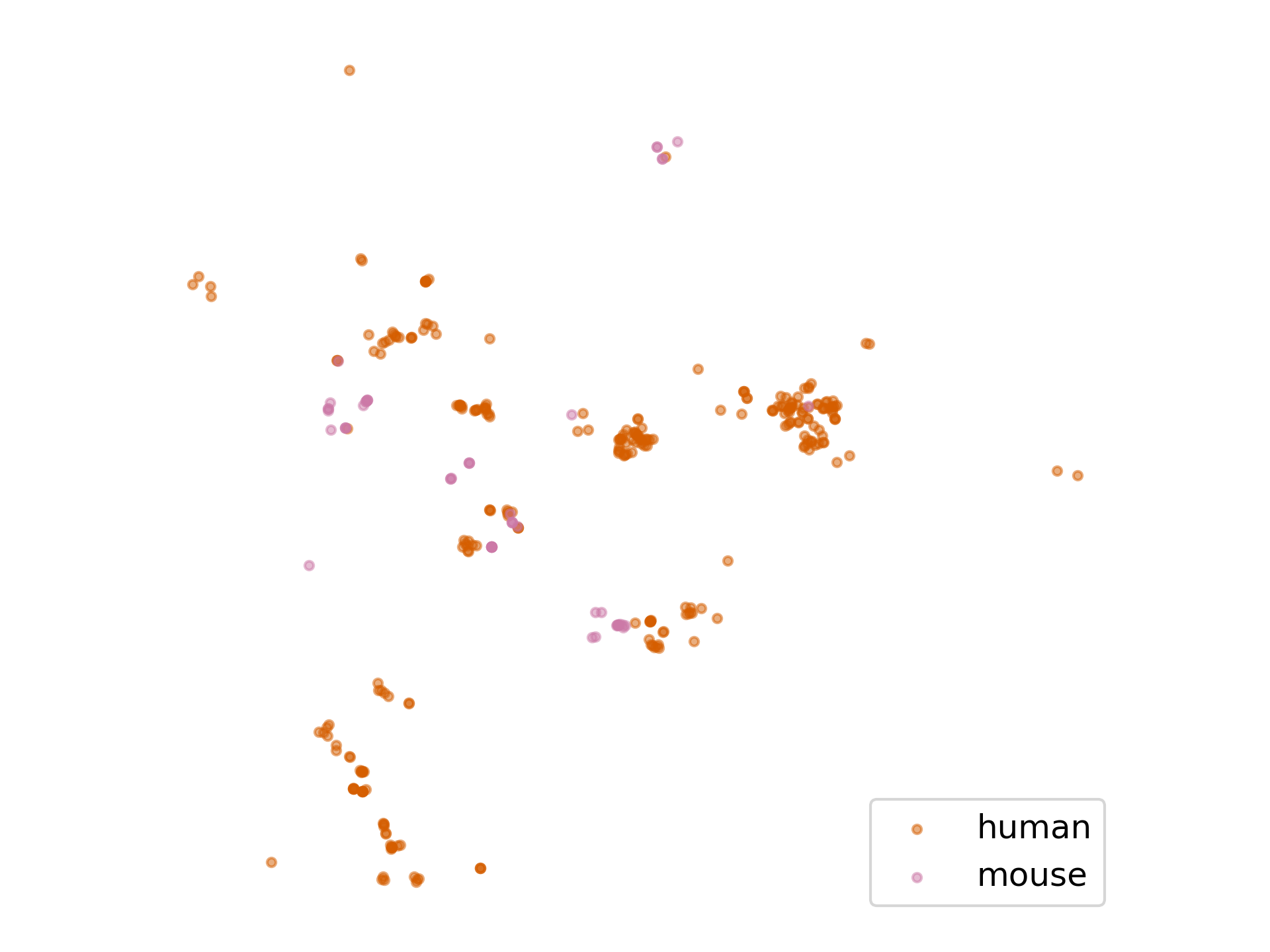}
  \caption{t-SNE plot of antibody embeddings colorized by ANARCI annotated species}
\end{figure}

\subsubsection{Zero-shot Prediction from Paired Antibody Perplexity}
Trained on clustered sequences, our model performs more weakly (p-value $<$ 0.05) on one dataset. Results are largely unaffected by sequence clustering.

\begin{figure}[H]
  \centering
  \includegraphics[width=0.75\linewidth]{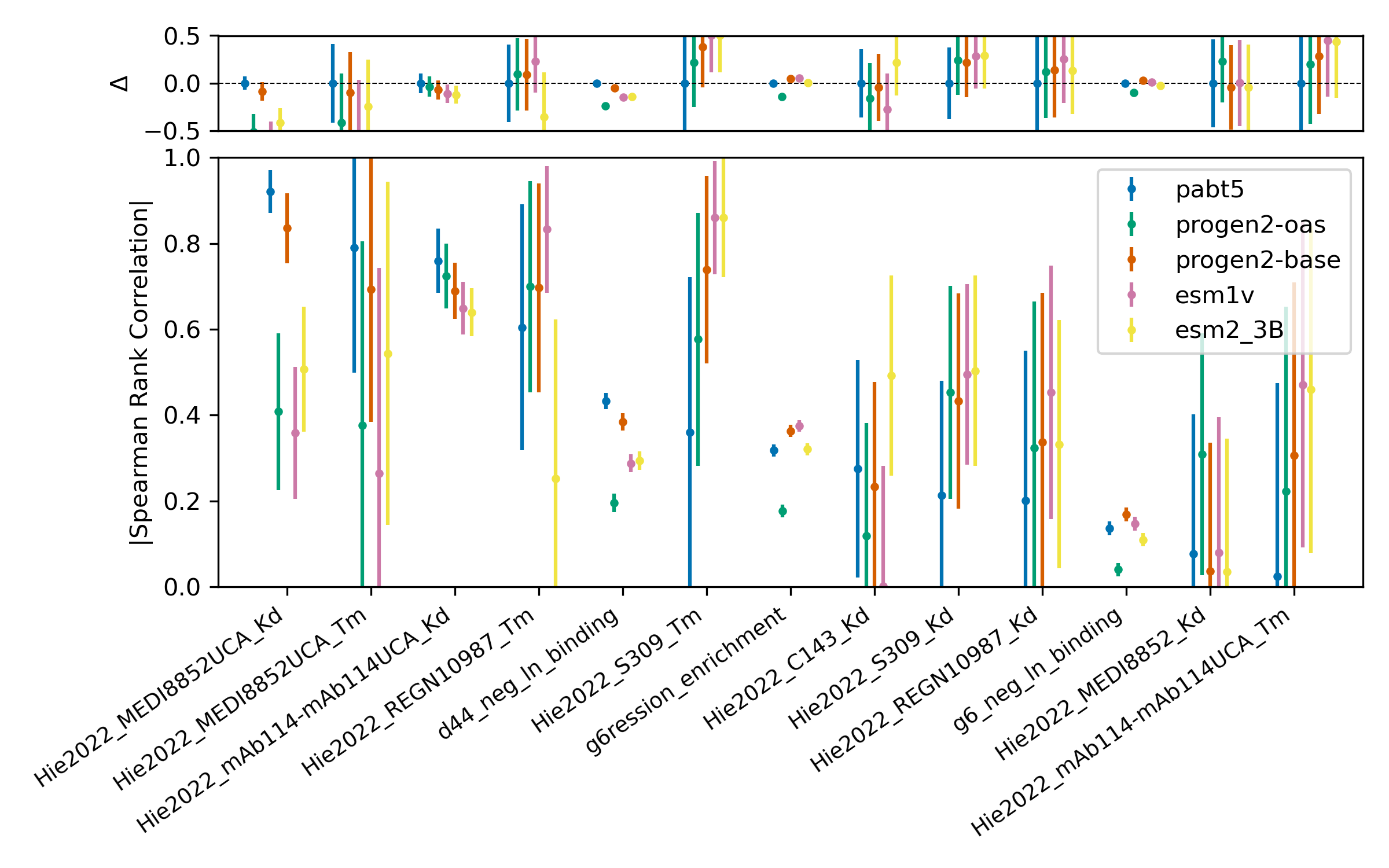}
  \caption{Zero-shot prediction performance on antibody measurements of our model and state-of-the-art on all datasets. x-axis represents datasets. (Top) The difference in absolute spearman rank correlation (SRC) between our model and state-of-the-art. (Bottom) Absolute SRC between model (pseudo-)perplexity and measurements. Error bars are estimated in standard deviation with 1000 bootstrap samples.}
  \label{fig:sota_zeroshot_linclust}
\end{figure}

\begin{figure}[H]
  \centering
  \includegraphics[width=0.75\linewidth]{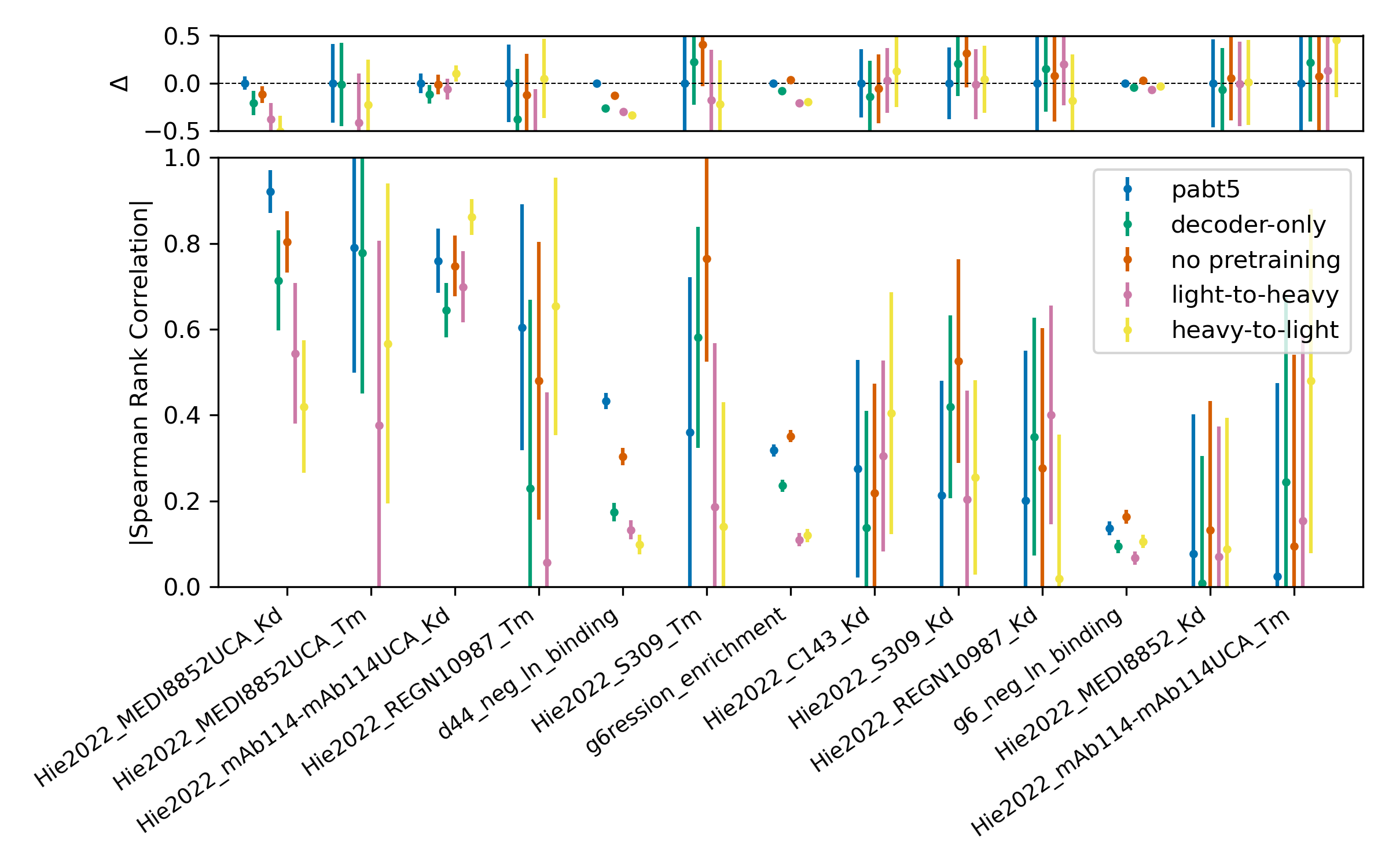}
  \caption{Ablation study on zero-shot prediction on all datasets. x-axis represents datasets. (Top) The difference in absolute spearman rank correlation (SRC) between our model and ablation. (Bottom) Absolute SRC between model (pseudo-)perplexity and measurements. Error bars are estimated in standard deviation with 1000 bootstrap samples.}
\end{figure}


\end{document}